\documentclass[journal]{IEEEtran}
 \pdfoutput=1
 \usepackage{cite}
\usepackage{amsmath,amssymb,amsfonts}
\usepackage{algorithmic}
\usepackage{graphicx}
\usepackage{textcomp}
\usepackage{framed,multirow}
\usepackage{latexsym}
\usepackage{amsmath}
\usepackage{threeparttable}
\usepackage{float}
\usepackage{array}
\usepackage{hhline}
\usepackage{colortbl}
\usepackage{stfloats}
\usepackage{booktabs}
\usepackage{multirow}
\usepackage{enumerate}
\usepackage{diagbox}
\usepackage{url}
\usepackage{hyperref}
\usepackage{makecell,rotating}
\usepackage{graphicx}
\usepackage{textcomp}
\usepackage{xcolor}
\usepackage{color}
\usepackage{tikz}

\DeclareGraphicsExtensions{.pdf,.jpeg,.png,.jpg}

\graphicspath{{paper_figures/}}
\newcommand{\tabincell}[2]{\begin{tabular}{@{}#1@{}}#2\end{tabular}}

\definecolor{mygray}{gray}{.9}

\definecolor{newcolor}{rgb}{.8,.349,.1}

\makeatletter

\newcommand{\Rmnum}[1]{\expandafter\@slowromancap\romannumeral #1@}
\makeatother

\title{ RestainNet: a self-supervised digital re-stainer for stain normalization}

\author{Bingchao Zhao, Jiatai Lin, Changhong Liang, Zongjian Yi, Xin Chen, Bingbing Li, Weihao Qiu, Danyi Li, Li Liang, Chu Han, and Zaiyi Liu
\thanks{This work was supported by the National Key R\&D Program of China 2017YFC1309100 and 2017YFC1309002, the National Science Fund for Distinguished Young Scholars 81925023, National Natural Science Foundation of China 81771912, 82071892, and 82072090, High-level Hospital Construction Project DFJH201805 and R\&D plan of Key Fields in Guangzhou (202007040001). (Corresponding authors: C Han: Z Liu.)}
\thanks{Bingchao Zhao is with the School of Medicine, South China University of Technology, 
	Guangzhou 510006, China, and with the Department of Radiology, Guangdong Provincial People's Hospital, Guangdong Academy of Medical Sciences, 510080, China. (e-mail: zbcajj@qq.com)}
\thanks{Jiatai Lin and Zongjian Yi are with the South China University of Technology, Guangzhou 510006, China, and with the Department of Radiology, Guangdong Provincial People's Hospital, Guangdong Academy of Medical Sciences, 510080, China (e-mail: linjiatai\_cs@163.com; jiangiegie@qq.com)}
\thanks{Bingbing Li is with Guangdong Provicial People's Hospital Ganzhou Hospital and Department of Pathology, Nanfang Hospital and Basic Medical College, Southern Medical University (email: libingbing19932021@126.com)}
\thanks{Li Liang, Weihao Qiu, Danyi Li are with Department of Pathology, Nanfang Hospital and Basic Medical College, Southern Medical University, Guangzhou 510515, Guangdong, China and Guangdong Province Key Laboratory of Molecular Tumor Pathology, Guangzhou 510515, Guangdong, China. (email: lli@smu.edu.cn; qiuweihao42@163.com; lidanyi26@163.com)}
\thanks{Xin Chen is with the Department of Radiology, Guangzhou First People's Hospital, School of
	Medicine, South China University of Technology, Guangzhou, 510180, China. (e-mail: wolfchenxin@163.com)}
\thanks{Chu Han is with 
	the Department of Radiology, Guangdong Provincial People's Hospital, Guangdong Academy of Medical Sciences, 510080, China. (e-mail: hanchu@gdph.org.cn)}
\thanks{Zaiyi Liu and Changhong Liang are with  
	the Department of Radiology, Guangdong Provincial People's Hospital, Guangdong Academy of Medical Sciences, 510080, China, and with the School of Medicine, South China University of Technology, Guangzhou 510006, China. (e-mail: zyliu@163.com; liangchanghong@gdph.org.cn)}}

\begin{document}
\maketitle

\IEEEtitleabstractindextext{\begin{abstract}
Color inconsistency is an inevitable challenge in computational pathology, which generally happens because of stain intensity variations or sections scanned by different scanners. It harms the pathological image analysis methods, especially the learning-based models. A series of approaches have been proposed for stain normalization. However, most of them are lack of flexibility in practice. In this paper, we formulated stain normalization as a digital re-staining process and proposed a self-supervised learning model, which is called RestainNet. Our network is regarded as a digital re-stainer which learns how to re-stain an unstained (grayscale) image. Two digital stains, Hematoxylin ($\mathcal{H}$) and Eosin ($\mathcal{E}$), were extracted from the original image by Beer-Lambert's Law. We proposed a staining loss to maintain the correctness of stain intensity during the re-staining process. Thanks to the self-supervised nature, paired training samples are no longer necessary, which demonstrates great flexibility in practical usage. Our RestainNet outperforms existing approaches and achieves state-of-the-art performance with regard to color correctness and structure preservation. We further conducted experiments on the segmentation and classification tasks and the proposed RestainNet achieved outstanding performance compared with SOTA methods. The self-supervised design allows the network learn any staining style with no extra effort.
\end{abstract}
\begin{IEEEkeywords}
Computational Pathology, \and Stain Normalization, \and Self-supervised Learning
\end{IEEEkeywords}}

\maketitle
\IEEEdisplaynontitleabstractindextext

\IEEEpeerreviewmaketitle

\section{Introduction}
Computational pathology is now becoming more and more popular with the invention of high-precision microscope section scanners and the development of deep neural networks. It aims to quantitatively evaluate the microenvironment of diseases in the microscopic level in an objective manner, and finally achieves precision medicine, i.e., disease diagnosis~\cite{yu2016predicting}, treatment response~\cite{2020Predicting}, and even gene prediction~\cite{2020Integrating}. Although CNN models have demonstrated outstanding performance in various computational pathology applications, most of them still suffer from color inconsistency problem.
\begin{figure}[t]
	\centering
	\setlength{\tabcolsep}{1pt}
	\begin{tabular}{cccc}
		\scriptsize{\rotatebox{90}{Aperio}}&
		\includegraphics[width=.30\linewidth]{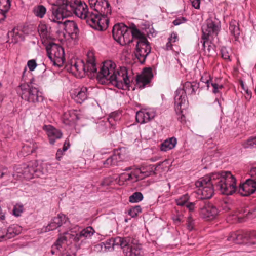}&
		\includegraphics[width=.30\linewidth]{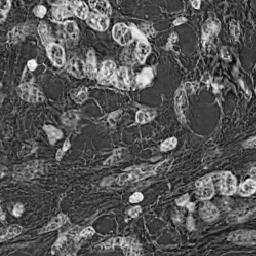}&
		\includegraphics[width=.30\linewidth]{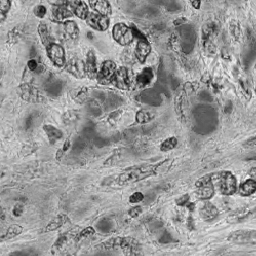}\\
		\scriptsize{\rotatebox{90}{Hamamatsu}}&
		\includegraphics[width=.30\linewidth]{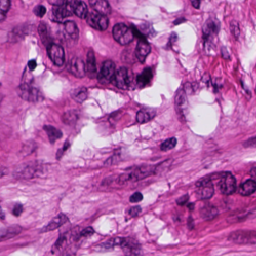}&
		\includegraphics[width=.30\linewidth]{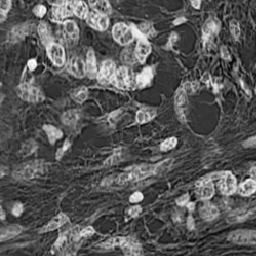}&
		\includegraphics[width=.30\linewidth]{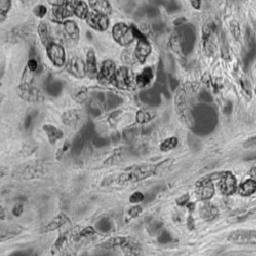}\\
		&(a) Original & (b) $\mathcal{H}$ & (c) $\mathcal{E}$
	\end{tabular}
	\caption{Two extracted digital dyes. (a) are two images scanned by two difference scanners, Aperio and Hamamatsu. (b) and (c) are two images extracted by Beer-Lambert's Law~\cite{parson2007modern}. They demonstrate consistent dyes intensity against different scanners.}
	\label{fig:stain_compare}
\end{figure}
It is introduced by the various steps during sectioning, i.e., inconsistent tissue thickness, different manufacturers of the dye, different sectioning protocols and etc~\cite{roy2018study}.
In addition, scanning with different scanners may also lead to color inconsistency problem, which harms the generalization of the learning-based models, especially when applying them to the data from different institutions.
That is the reason why color normalization is always an essential preprocessing step before model training~\cite{graham2019hover,2017The}.

To alleviate the color inconsistency problem, many color normalization methods have been proposed. Traditional methods normalize pathological images in a statistical manner~\cite{jain1989fundamentals,reinhard2001color}.
Computational-based methods take characteristics of H\&E stained images into consideration, and decompose them into different dye intensity images~\cite{Ruifrok2001Quantification,2009A,2013Blind,vahadane2016structure,magee2009colour}. 
And then they approximate the color distribution of the source image to the distribution of a selected template image. However, the selection bias of the reference images may lead to inconsistent results. Moreover, optimization-based approaches are easily trapped in local optimum~\cite{bentaieb2017adversarial}.
Recently, GAN-based models~\cite{Shaban2019Staingan,salehi2020pix2pix} have been proposed and achieved promising results. Due to the data-driven nature, they require a large amount of paired training samples, which are hard to collect. A feasible solution is to scan the pathological section using different scanners. But we still cannot guarantee perfect groundtruth images because of the misalignment problem. Additional alignment processes may damage the structure of the original image. Moreover, it is impractical to re-collect the training data when an unseen staining style comes. Besides the stain normalization performance, the flexibility of the model extension should be also considered.

In this paper, we formulate stain normalization as a digital re-staining process. We first de-stain the pathological image by transforming the input image into a grayscale image. In order to preserve structure, $L$ channel is extracted from $Lab$ color space. Then we re-stain it by applying two digital dyes, $\mathcal{H}$ and $\mathcal{E}$. According to the fact that hematoxylin always stains nuclei blue and Eosin always stains the cytoplasm and extracellular matrix pink, we extract two digital dyes $\mathcal{H}$ and $\mathcal{E}$ by Beer-Lambert’s Law~\cite{parson2007modern}. It can be observed that $\mathcal{H}$ and $\mathcal{E}$ are insensitive to different stain styles as demonstrated in Fig.~\ref{fig:stain_compare}.

To achieve this, we proposed a self-supervised learning model, called RestainNet, which plays a role as a digital re-stainer. The network learns how to stain an unstained (grayscale) image with two digital stains $\mathcal{H}$ and $\mathcal{E}$. In order to maintain the correctness of digital stains after re-staining process, we proposed a new staining loss. GAN loss is introduced to differentiate the real or the generated images. Due to the self-supervised manner, we do not have to collect paired images for network training which is easily implemented into practical applications. Furthermore, our model is able to generate visually pleasant normalization results while preserving the structure information. Several experiments were conducted on public datasets to evaluate the effectiveness of stain normalization results. Our proposed model achieves state-of-the-art performance in all the quantitative measurements compared with existing approaches. We also conducted experiments to prove that our proposed model can benefit the following segmentation and classification tasks in practice. Furthermore, the ablation study validates the stability of the model with regard to the vibration of the digital dyes.

Our major contributions can be summarized as follows:
\begin{itemize}
	\item We proposed a self-supervised stain normalization model, called RestainNet. Thanks to the self-supervised nature, paired training images are no longer necessary, which shows great flexibility in practical usage.
	\item We proposed a new staining loss to guarantee the correctness of digital dyes.
	\item Our RestainNet achieves state-of-the-art performance comparing with existing models in three different tasks, the color normalization task, the segmentation task and the classification task.
\end{itemize}
\section{Related Works}
\label{relate_works}
\subsection{Computational-based Normalization}
To reduce the impact of color inconsistency, some works convert RGB images into grayscale images  to protect the image texture information~\cite{gurcan2009histopathological}. 
Although these methods obtain stable texture features that do not change with the stain style, but discard the color features that contain richer semantic information. 
If we regard each stain style as specific color distribution, an intuitive method is to convert all images into the same color distribution space. 
This conversion is usually not performed in the RGB space, because it is more difficult to normalize the type of color than the intensity. 
Reinhard~et~al.~\cite{reinhard2001color} use the de-correlated Lab components obtained from the RGB image to perform statistical matching with the histogram of the target image. This method may result in an image that does not conform to the actual distribution because the types and proportions of tissue components in most images are not constant. 

The color of the pathological image is characterized by the dye and its concentration attached to the tissue. The RGB image can be converted to the dye concentration space through the specified mapping method, providing a more reasonable paradigm for describing the color distribution of pathological images. Among them, the widely used color deconvolution algorithm~\cite{Ruifrok2001Quantification} decomposes the RGB image into Hematoxylin (H), Eosin (E), and Diaminobenzidine (DAB) components through the "staining matrix" obtained by statistical evaluation. The data-dependent "staining matrix" may lead to miscalculations of dye concentration when generalized to unseen data sets. Vahadane~et~al.~\cite{vahadane2016structure} perform stain separation based on non-negative matrix factorization which,  in a supervised way, clusters the image to two empirically selected tissue clusters. 
Furthermore, Janowczyk~et~al.~\cite{janowczyk2017stain} use Sparse AutoEncoders (StaNoSA) to independently performed normalization of each cluster to alleviate the impact of the tissue clusters imbalance among images.

These Computational-based algorithms perform stain normalization by mapping the image to the color space represented by the target image or selected distribution. But this characteristic makes them limited in actual implementation, because the normalized image may appear with tissue categories outside the selected image.

\begin{figure*}[htp]
	\centering
	\includegraphics[width=\linewidth]{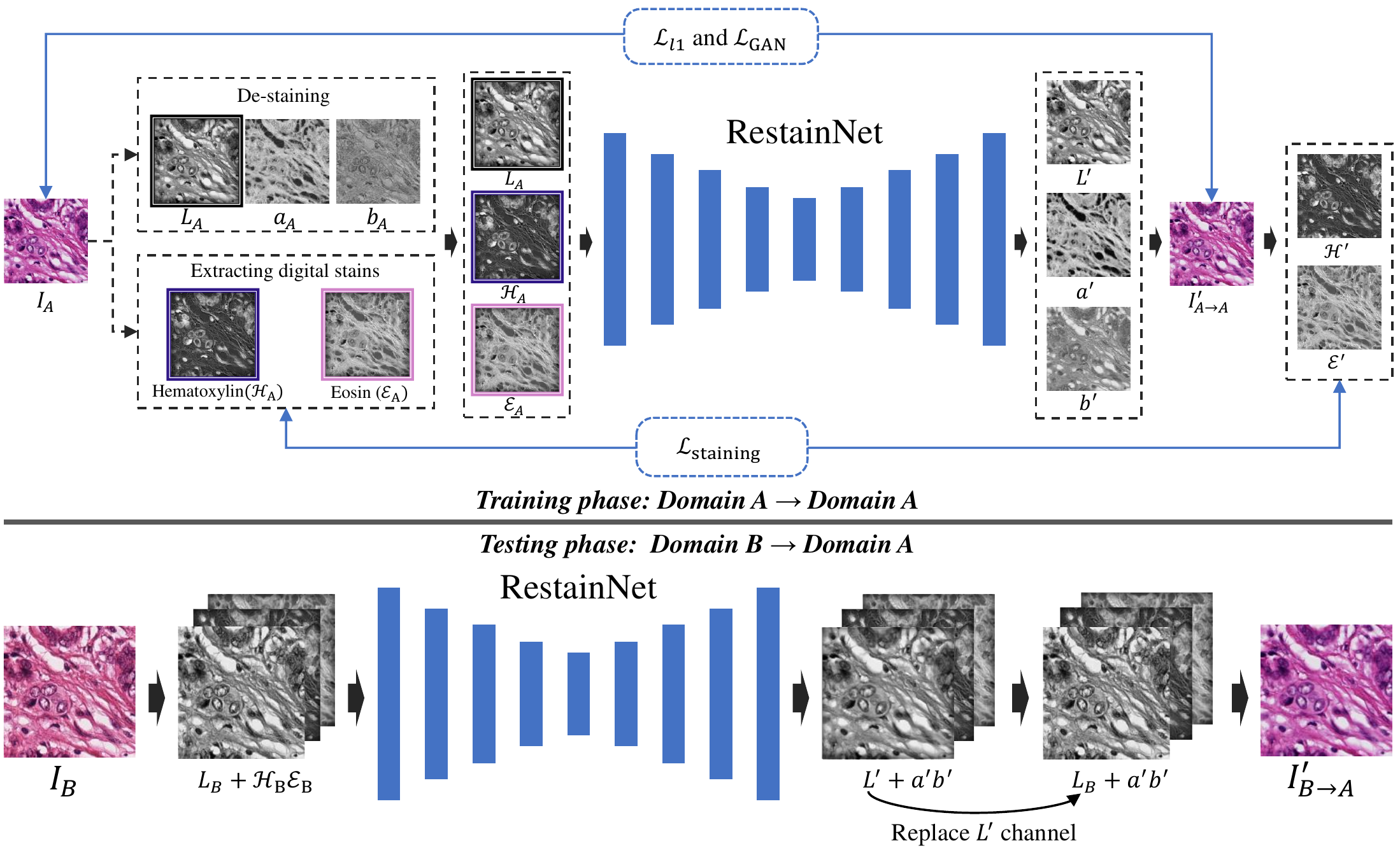}
	\caption{Overview of RestainNet. $L$ is the luminance channel of $Lab$ color space. $\mathcal{H}$ and $\mathcal{E}$ are two digital dyes extracted by Beer-Lambert's Law~\cite{parson2007modern}. During the training phase, RestainNet learns how to re-stain the grayscale image back to the original image in a self-supervised manner. During the testing phase, images from the source domain $B$ can be transferred to the target domain $A$.}
	\label{fig:network}
\end{figure*}
\subsection{Learning-based Normalization}
With the rapid progress of deep learning technology, convolutional neural networks have been widely applied in stain normalization tasks. Bug~et~al.~\cite{bug2017context} proposed a feature-aware framework that formulates the normalization task as a non-linear mapping between pixels but causes image compression. 
Inspired by the promising performance of the generative model (e.g., GAN~\cite{mirza2014conditional}, CycleGANs~\cite{2017Unpaired}), many works regard stain normalization as an image style transfer task that aims to synthesize the target image texture but preserves the semantic content of the source image.
Shaban~et~al.~\cite{Shaban2019Staingan} introduced CycleGANs~\cite{2017Unpaired} to realize the style transfer between images from two different scanners. 
Salehi~et~al.~\cite{salehi2020pix2pix} designed an unsupervised model based on conditional generative adversarial networks (cGANs)~\cite{mirza2014conditional} by coloring grayscale images to the target stain style. 
A multi-task network equipped with normalized branches and task-specific branches~\cite{bentaieb2017adversarial} is proposed to improve the generalization ability of target tasks on multiple data sets.
Liang~et~al.\cite{liang2020stain} trained a multi-task network that combined clinical diagnosis information to synthesize the structure-maintained normalized image.

The learning-based method trains the huge parameters using a well-designed learning strategy to transfers the source image to the target style. 
With a large amount of training data, the deep learning model can fit the color distribution of a specified dataset, and generate images with richer colors than the traditional mathematical method that refers to a single target image.
The model synthesis images with richer semantic information, but the large number of high-quality images required for training restricts its practicality.
Moreover, existing methods transfer styles through image texture similarity but rarely use prior knowledge, which may cause normalization errors when feeding images with different tissue types or proportions from the training images.

\section{Methodology}

In this section, we introduce the proposed model RestainNet for stain normalization. Section~\ref{sec:formulation} shows the problem formulation. Section~\ref{sec:restainnet} demonstrates the details of the complete model. Section~\ref{sec:loss} shows the loss function of the model.
\subsection{Problem Formulation}
\label{sec:formulation}
In this paper, we proposed a simple and flexible model for stain normalization. Generally, we formulate stain normalization as a digital re-staining process, as demonstrated in Fig.~\ref{fig:network}. Given an input image, we first de-stain it by extracting the grayscale image. To preserve the structure of the original image, $L$ channel is extracted from $Lab$ color space. Then two digital dyes, Hematoxylin ($\mathcal{H}$) and Eosin ($\mathcal{E}$), are extracted to represent the dye intensity maps of each pixel. 

Given a target domain $\mathcal{D}_A$ and a source domain $\mathcal{D}_B$, let us denote two images from two different domains as $I_A\in\mathcal{D}_A$ and $I_B\in\mathcal{D}_B$. $f_{\theta}(\cdot)$ denotes the RestainNet with the parameter set $\theta$.
\begin{itemize}
	\item In \textit{training phase}, we aim to learn the internal characteristics of the target domain $\mathcal{D}_A$ in a self-supervised manner. Given the grayscale image $L_A$ and two digital dyes $\mathcal{H}_A$ and $\mathcal{E}_A$, RestainNet generates the re-stained image $I'_{A\rightarrow A}$ by $f_{\theta}(L_A,\mathcal{H}_A,\mathcal{E}_A)$ which satisfies $I'_{A\rightarrow A} \approx I_A$.
	
	\item In \textit{testing phase}, RestainNet allows input images from any other source domain, e.g. domain $\mathcal{D}_B$, and re-stains it to the target domain $\mathcal{D}_A$, i.e., $f_{\theta}(L_B,\mathcal{H}_B,\mathcal{E}_B) = I'_{B\rightarrow A}$. Finally, we replace $(L'_A,\mathcal{H}_A,\mathcal{E}_A)$ with $(L_B,\mathcal{H}_A,\mathcal{E}_A)$ to save the complete texture information. In this paper, the results of the proposed model are obtained in this way.
\end{itemize}

\subsection{RestainNet}
\label{sec:restainnet}
\textbf{Network architecture:} RestainNet intends to learn to stain the de-stained image with the guidance of the H and E digital dyes through a self-supervised learning training phase. 
The re-stained images should maintain the original resolution with the input, to achieve a structure-preserving and practical normalizationing
For this reason, the re-stainer can be generalized to a structure with such properties, and not being restricted to a fixed network. 
The U-net~\cite{2015U} structure maintains the invariance of the output resolution through the same number of upsampling and downsampling phases and employees skip links to protect the high-frequency information, which can serve as an eligible re-stainer.
In addition, the autoencoder-decoder structure with a limited number of encoding can also achieve the above requirements, otherwise, the information loss caused by excessive downsampling can result in the inability to generate a complete image.  
We find that an autoencoder with two downsamplings and two upsampling can achieve the above requirements. 
In this paper, a U-net structure is adopted, although these two structures have shown similar performance in the experiment.

\textbf{Digital stains extraction:} Hematoxylin, and Eosin are two stains that make pathological section pigment for better observation. During the staining process, hematoxylin is used for staining the cells into blue and eosin is used for staining the cytoplasm into pink. Therefore, the color appearance of the H\&E stained pathological section is determined by the specific dye attached to the tissue.

We want to extract two digital dyes for our proposed RestainNet. According to Beer-Lambert’s Law~\cite{parson2007modern}, the transmission of light through a material is described by the incident light $\mathcal{I}_{0}$, the dye intensity $A$ and the absorption factor $c$ as follows:
\begin{equation}\label{eq:Lambert-Beer} 
	\mathcal{I}_{r} = \mathcal{I}_{0}exp(-Ac)
\end{equation}
where $\mathcal{I}_{r}$ is the received light passing through the section. After a simple mathematical transformation of Equation~\ref{eq:Lambert-Beer}, we have:
\begin{equation}\label{eq:OD}
	OD = -\log\frac{\mathcal{I}_{r}}{\mathcal{I}_{0}} = Ac
\end{equation}
Note that $OD$ is the abbreviation of optical density. The dye intensity $A$ can be obtained by decomposing the optical intensity $OD$ due to the linearity. Through experimental statistics, we can get a normalized pure stain OD matrix M:

$$M = \left[
\begin{array}{ccc}
	0.65 &	0.7 & 0.29 \\
	0.07 & 0.99 & 0.11\\
	0.27 & 0.57 & 0.78
\end{array} \right]
$$
On this basis, we can further calculate the optical density $y$ of each pixel by the following equation:
\begin{equation}\label{eq:stain_intensity}
	y = AM \Rightarrow A=yM^{-1}
\end{equation}


\subsection{Loss function}
\label{sec:loss}
We first introduce the adversarial loss to evaluate the re-stained image.
$$\mathcal{L}_{GAN} = E_{I \sim p_{data}(I)}[\log D(I)] +  E_{I' \sim p_{z}({I'})}[log(1 - D(G({I'}))]$$
where $I$ is the input and $I'$ is the generated image, model $C$ maximize the input while $G$ minimizes the $log(1 - D(G({I'}))$.

Since the model generates images in $Lab$ color space. L1 loss is applied on $Lab$ color space as follows:
\begin{equation}
	\mathcal{L}_{l1} = \Sigma(||L-L'||_{1}+||a-a'||_{1}+||b-b'||_{1})
\end{equation}

The existing learning-based normalization algorithms usually evaluate the generated images in the RGB color space, but ignore the unique optical characteristics of pathological images. To preserve the consistency of the digital stain intensity during the re-staining process, we proposed a new loss, called staining loss. From the re-stained image $I'$ generated by RestainNet, we extract its digital dyes $\mathcal{H}'$ and $\mathcal{E}'$ and compare them with the original ones, $\mathcal{H}$ and $\mathcal{E}$. And the staining loss is calculated by measuring the L1 distance of them.

\begin{equation}
	\mathcal{L}_{staining} = \Sigma(||\mathcal{H} - \mathcal{H^{'}}||_{1}+||\mathcal{E} - \mathcal{E^{'}}||_{1})
\end{equation}

The final objective function is defined as:
\begin{equation}\label{eq:total_loss}
	\mathcal{L} = \lambda_{GAN}\mathcal{L}_{GAN} + \lambda_{l1}\mathcal{L}_{l1} + \lambda_{staining}\mathcal{L}_{staining}
\end{equation}
where $\lambda_{\{GAN, l1, staining\}}$ are the hyperparameters.

\subsection{Implementation and Training Details}
We build our model under Ubuntu18.04 using Pytorch1.8 architecture. The optimizer is Adam~\cite{kingma2014adam}. The momentum parameter $\beta$ is 0.5. The learning rate is 0.0002 with an exponential decay of 0.6. The network runs on a workstation with an NVIDIA GeForce RTX 3090. We select the best performing model on the validation set as our final model. Hyperparameters $\lambda_{GAN}$, $\lambda_{l1}$, $\lambda_{stain}$ are set to 0.1, 1, 1, respectively.
\section{Experiments}
\begin{figure*}[t]
	\setlength{\tabcolsep}{1pt}
	\begin{tabular}{lccccccc}
		A&
		\begin{tikzpicture}[
			every node/.style={anchor=south east,inner sep=0pt},
			x=.1mm, y=.1mm,
			]
			\node (fig1) at (0,0)
			{\includegraphics[width=.155\linewidth]{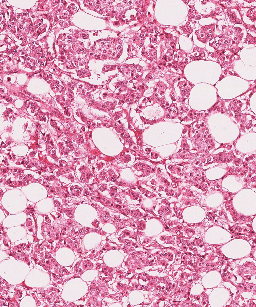}};
			\node (fig2) at (0,0)
			{\fcolorbox{blue}{blue}{\includegraphics[width=0.06\linewidth]{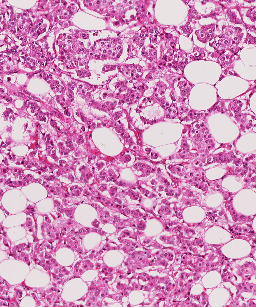}}};
		\end{tikzpicture}&
		\includegraphics[width=.155\linewidth]{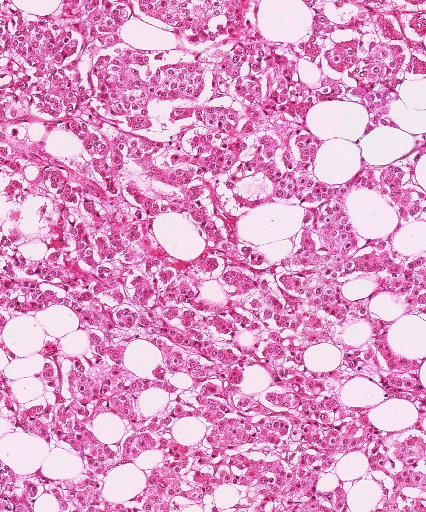}&
		\includegraphics[width=.155\linewidth]{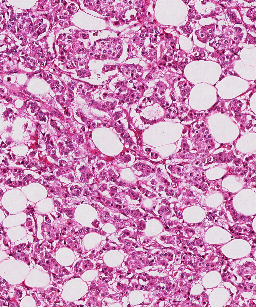}&
		\includegraphics[width=.155\linewidth]{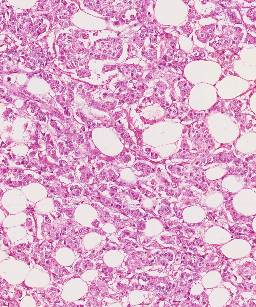}&
		\includegraphics[width=.155\linewidth]{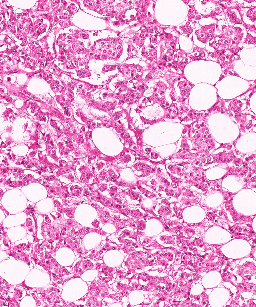}&
		\includegraphics[width=.155\linewidth]{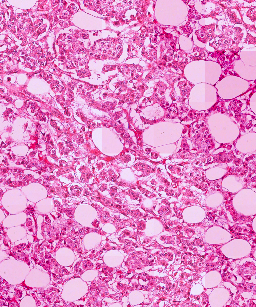} \\
		
		B&
		\begin{tikzpicture}[
			every node/.style={anchor=south east,inner sep=0pt},
			x=.1mm, y=.1mm,
			]
			\node (fig1) at (0,0)
			{\includegraphics[width=.155\linewidth]{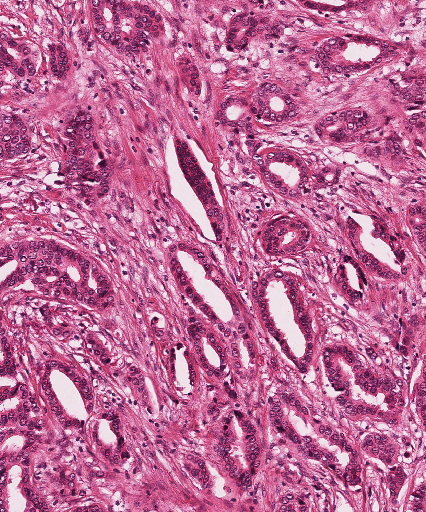}};
			\node (fig2) at (0,0)
			{\fcolorbox{blue}{blue}{\includegraphics[width=0.06\linewidth]{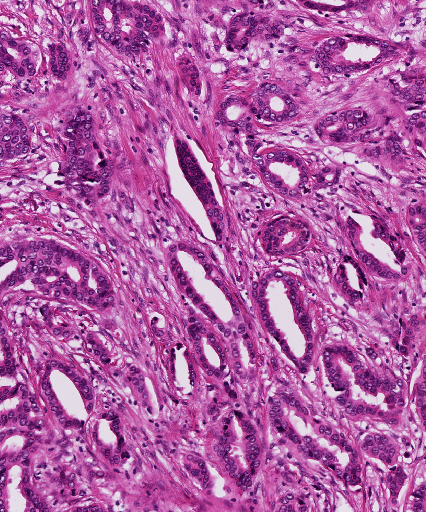}}};
		\end{tikzpicture}&
		\includegraphics[width=.155\linewidth]{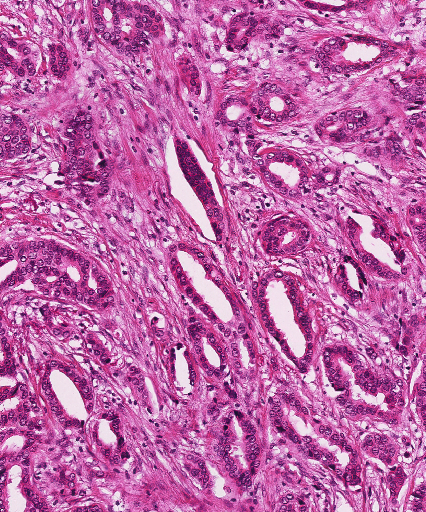}&
		\includegraphics[width=.155\linewidth]{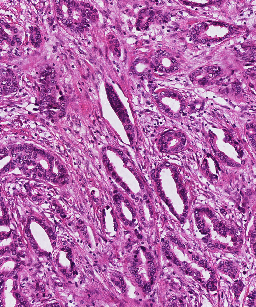} &
		\includegraphics[width=.155\linewidth]{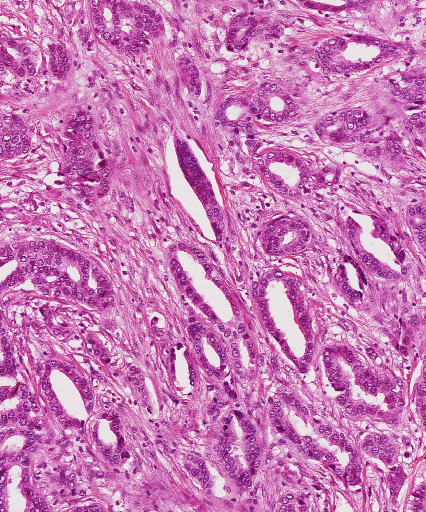}&
		\includegraphics[width=.155\linewidth]{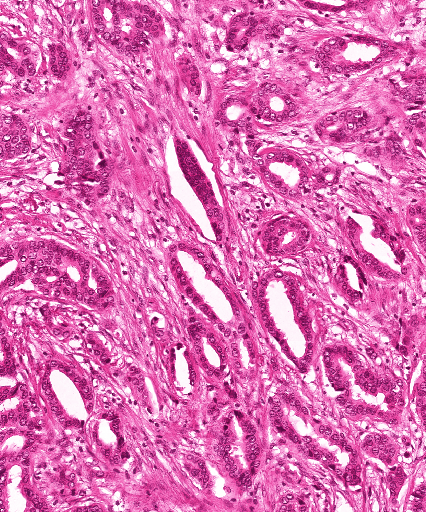}&
		\includegraphics[width=.155\linewidth]{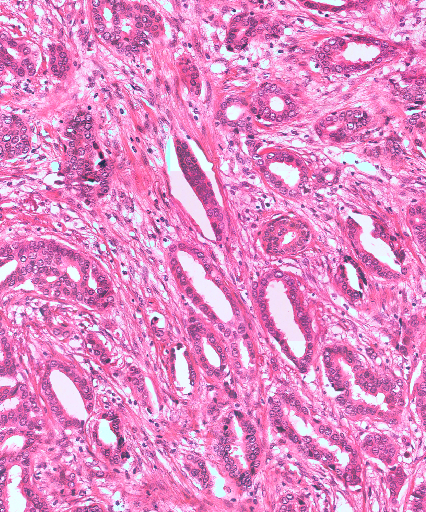}\\
		
		C&
		\includegraphics[width=.155\linewidth]{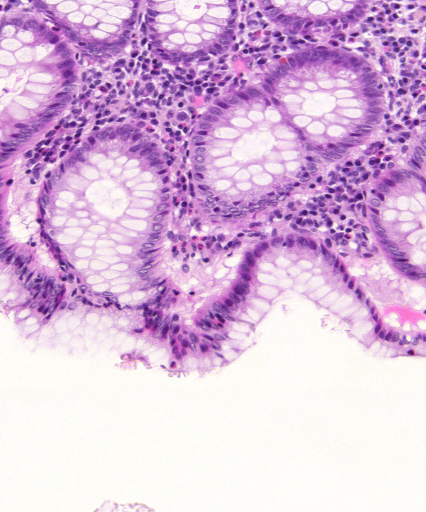}& 
		\includegraphics[width=.155\linewidth]{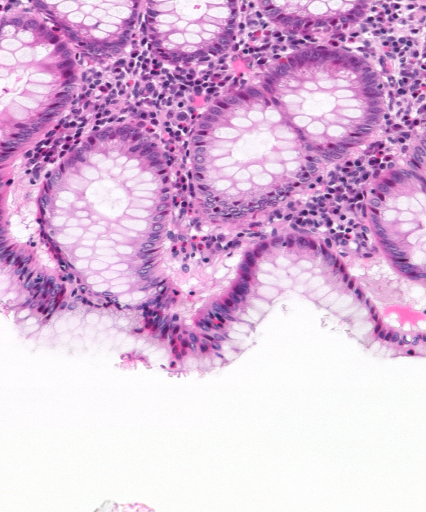}& 
		\includegraphics[width=.155\linewidth]{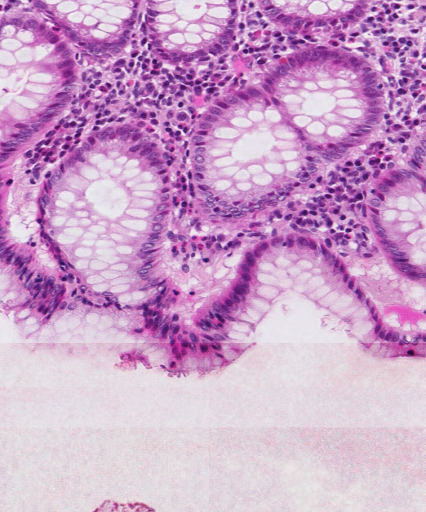}&
		\includegraphics[width=.155\linewidth]{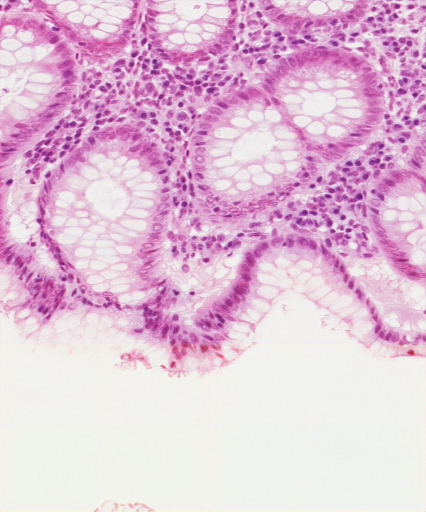}&
		\includegraphics[width=.155\linewidth]{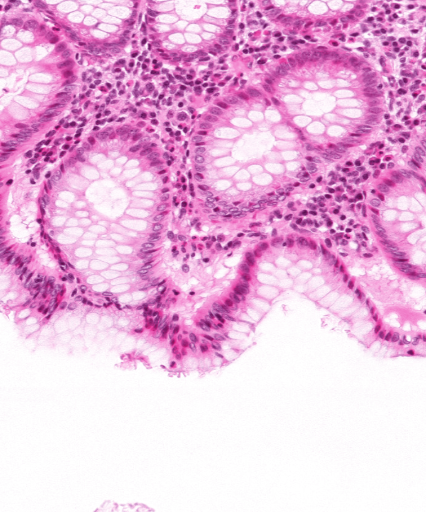}& 
		\includegraphics[width=.155\linewidth]{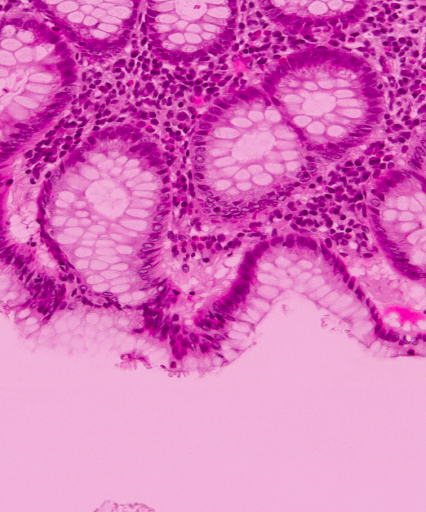}\\
		
		D&
		\includegraphics[width=.155\linewidth]{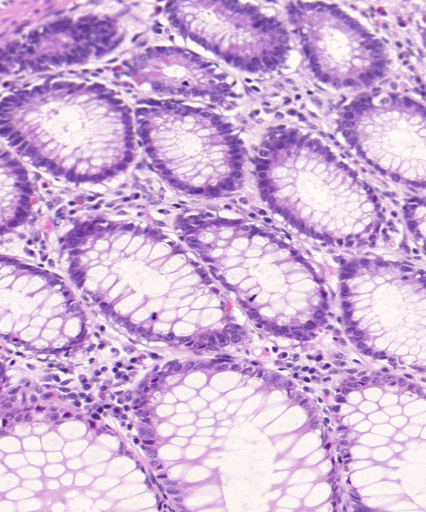}& 
		\includegraphics[width=.155\linewidth]{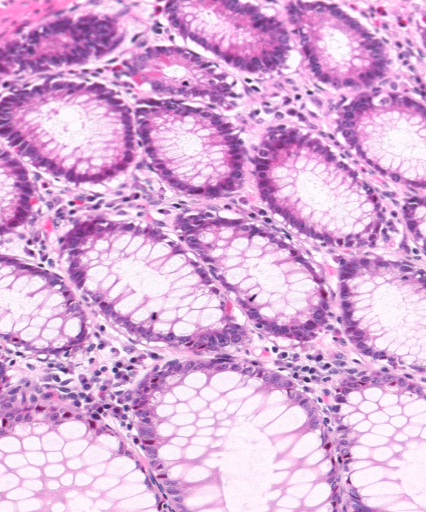}& 
		\includegraphics[width=.155\linewidth]{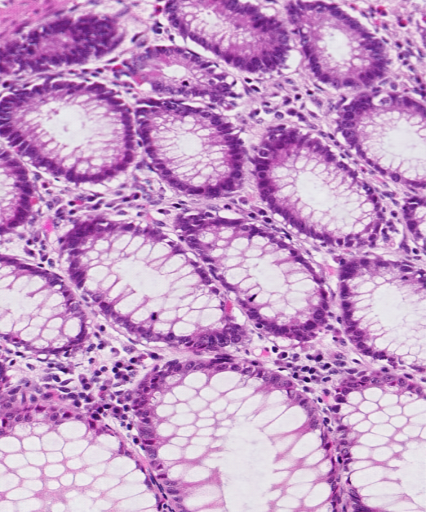}&
		\includegraphics[width=.155\linewidth]{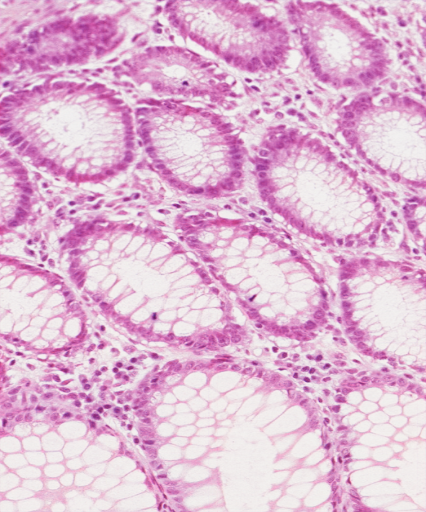}&
		\includegraphics[width=.155\linewidth]{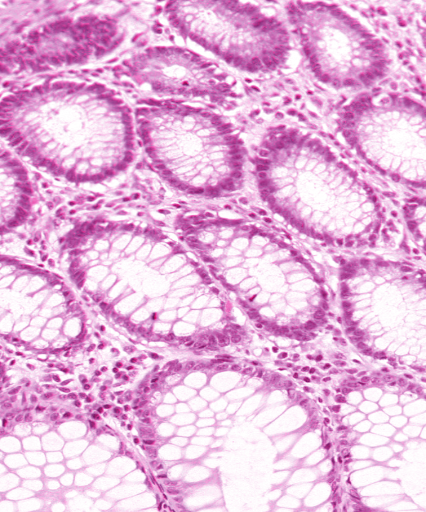}& 
		\includegraphics[width=.155\linewidth]{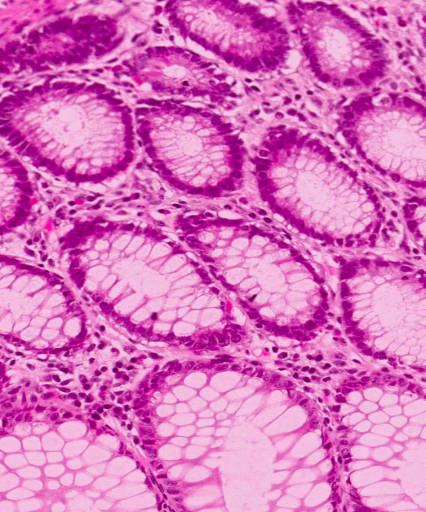}\\
		
		&(a) Input& (b) Ours  & (c) StainGAN~\cite{Shaban2019Staingan} & (d) STST~\cite{salehi2020pix2pix} & (e) Vahadane~\cite{vahadane2016structure} & (f) Reinhard~\cite{reinhard2001color} 
	\end{tabular}
	\caption{Qualitative comparison with different methods. (a) is the input images scanned by Aperio (source domain). Images in blue box were scanned by Hamamatsu (target domain). (b)-(f) are the results generated from different methods. Since the input images in row C and D are from MICCAI’16 GlaS challenge dataset, there is no reference image from the target domain.}
	\label{fig:vis_result}
\end{figure*}

\subsection{Dataset and Evaluation Metrics}
We evaluate our proposed method in three datasets, including a stain normalization dataset (MITOS-ATYPIA14 challenge~\footnote{https://mitos-atypia-14.grand-challenge.org/Home/}), a patch-level classification dataset (Zhao~et~al~\cite{0Artificial}) and a segmentation dataset (MICCAI’16 GlaS challenge dataset~\cite{2016Gland}).

MITOS-ATYPIA14 challenge contains images with three magnifications of 10X, 20X, and 40X, which were respectively scanned by two different scanners Aperio Scanscope XT and Hamamatsu Nanozoomer 2.0-HT. Our paper uses images under 20X magnification, which contains 424 images, including 300 images for the training set and 124 images for the testing set. All the images were cropped into 256$\times$256 without overlapping. We randomly selected 600 patches from the training set for validating our network. The validation set was not involved in network training. Due to the self-supervised nature, our model does not require paired images for training. So groundtruth images are only used for validating the performance of the proposed model.

Zhao~\cite{0Artificial} dataset, a colorectal cancer classification dataset, is committed to achieving tissue segmentation of pathological images through the patches classification task and consists of multiple centers and multiple tissues pathological images. The entire data set contains a total of nine tissues, which are blank areas and eight tissue types: adipose, debris, lymphocyte aggregates, mucus, muscle, normal mucosa, stroma, and tumor epithelium. The complete dataset consists of four subsets, which are training set (283.1k patches), test set1 (16.3k patches), test set (22.5k patches),  TSR evaluation set (126 blocks).
In this paper, we only use the training set and the test set1 and separate the data from the TCGA as the validation set.

MICCAI’16 GlaS challenge dataset~\cite{2016Gland} is published for the study of gland segmentation of colorectal cancer. It contains 165 $\mathcal{H}$\&$\mathcal{E}$ stained pathological images and corresponding gland segmentation masks. All images are divided into three subsets: Training Part (85 images), Test Part A (60 images), and Test Part B (20 images). In the experiment of this paper, all the images are cropped into 256 $\times$ 256 pixels patches, we use 30\% of the training part as the validation set and merge the two test parts. 

We used multiple quantitative measurements to evaluate the performance of our network, including: Feature Similarity (FSIM) Index~\cite{Zhang2011FSIM}, Peak Signal-to-Noise Ratio (PSNR), Erreur Relative Globale Adimensionnelle de Synthèse (ERGAS),Universal Quality Index (UQI)~\cite{2002A}, Structural Similarity Index (SSIM)~\cite{Wang2004Image}, Mean Squared Error (MSE),Relative Average Spectral Error (RASE), Root Mean Square Error (RMSE)~\cite{willmott2005advantages},  Multi-scale Structural Similarity Index (MS-SSIM)~\cite{2003Multiscale}.
\begin{table*}[t]
	\caption{Quantitative comparisons.}
	\begin{center}
		\scriptsize{
			\begin{tabular}{|c|c|c|c|c|c|c|}
				\hline
				Method &STST~\cite{salehi2020pix2pix} & Reinhard~\cite{reinhard2001color}& Vahadane~\cite{vahadane2016structure} & StainGAN~\cite{Shaban2019Staingan}& 
				 \tabincell{c}{Ours\\(w/o staining loss)} &\textbf{Ours} \\
				\hline
				
				FSIM 	&0.878 $\pm$ 0.052 	&0.872 $\pm$ 0.060 		&0.884 $\pm$ 0.053	&0.904 $\pm$ 0.023	&0.901 $\pm$ 0.033 &\textbf{0.917 $\pm$ 0.019}\\
				PSNR 	&18.627 $\pm$ 2.731 &19.384 $\pm$ 2.772 	&18.571 $\pm$ 2.864	&20.689 $\pm$ 3.422	&20.505 $\pm$ 3.570 &\textbf{21.550 $\pm$ 3.665}\\
				ERGAS 	&7.166 $\pm$ 2.75E3 &7.03E3 $\pm$ 2.55E3 	&7.39E3 $\pm$2.80E3	&5.91E3 $\pm$ 2.98E3&6.37E3 $\pm$3.52E3 &\textbf{5.70E3 $\pm$ 2.83E3}\\
				UQI 	&0.951 $\pm$ 0.043 	&0.964 $\pm$ 0.038 		&0.96 $\pm$ 0.045	&0.966 $\pm$ 0.056	&0.93 $\pm$ 0.065 &\textbf{0.973 $\pm$ 0.042}\\
				SSIM 	&0.842 $\pm$ 0.102 	&0.843 $\pm$ 0.05 		&0.863 $\pm$ 0.079	&0.916 $\pm$ 0.018	&0.873 $\pm$ 0.066 &\textbf{0.916 $\pm$ 0.018}\\
				MSE 	&1.19E3 $\pm$ 1.25E3 &0.95E3 $\pm$ 0.77E3 	&1.23E3 $\pm$ 1.30E3&0.85E3 $\pm$ 0.82E3&0.87E3 $\pm$ 1.105E3 &\textbf{0.76E3 $\pm$ 0.77E3}\\
				MS-SSIM &0.924 $\pm$ 0.066 	&0.941 $\pm$ 0.061 		&0.947 $\pm$ 0.061	&0.954 $\pm$ 0.024	&0.953 $\pm$ 0.031 &\textbf{0.976 $\pm$ 0.006}\\
				RASE 	&1.03E3 $\pm$ 3.92E2 &1.00E3 $\pm$ 3.65E2 	&1.05E3 $\pm$ 4.01E2&1.02E2 $\pm$ 4.02E2&0.917E3 $\pm$ 4.12E2 &\textbf{8.12E2 $\pm$ 4.10E2}\\
				RMSE 	&31.728 $\pm$ 13.415 &28.938 $\pm$ 10.582 	&32.096 $\pm$ 13.982&25.764 $\pm$ 14.12	&26.900 $\pm$ 14.818 &\textbf{23.770 $\pm$ 14.02}\\
				\hline
			\end{tabular}
		}
		\label{tab:result}
	\end{center}
\end{table*}
\subsection{Comparisons with Existing Approaches}
We use the same setting with the existing methods for a fair comparison. Images scanned by the Hamamatsu Nanozoomer 2.0-HT scanner are regarded as the target domain, and images scanned by the Aperio Scanscope XT scanner are regarded as the source domain.

Table~\ref{tab:result} demonstrates the quantitative comparisons with two computational-based stain normalization methods~\cite{reinhard2001color,vahadane2016structure} and two GAN-based models, StainGAN~\cite{Shaban2019Staingan} and STST~\cite{salehi2020pix2pix}.
Reinhard~et~al~\cite{reinhard2001color} match the histograms after converting the RGB image to the Lab color space. Vahadane~et~al~\cite{vahadane2016structure} perform stain separation base on non-negative matrix factorization to improve the accuracy of the dye concentration matrix obtained.
Since the proposed model performs stain normalization in $Lab$ color space, an unchanged $L$ channel helps to preserve image structure. Moreover, two digital dyes $\mathcal{H}$ and $\mathcal{E}$ demonstrate great consistency of dye concentration with different scanners. Therefore, our model demonstrates superiority in both structure-preserving and stain style transfer accuracy. We also evaluate the effectiveness of our proposed staining loss by comparing our model without this loss in Table~\ref{tab:result}. With the proposed staining loss, RestainNet gets overall improvements in all the quantitative measurements.

Fig.~\ref{fig:vis_result} shows the qualitative comparisons with the existing methods mentioned in Table~\ref{tab:result}. 
The computational-based methods (e.g. (e) and (f)) have good color consistency and are less affected by the source image, but the imbalance of tissue clusters still misleads the method to perform a global stain style transfer. For example in (f), the blank area is incorrectly stained by the algorithm to reduce the histogram distance between the source image and the target image.
In comparison, GAN-based methods (c) and (d) demonstrate great flexibility in global-local stain style transfer. However, StainGAN (c) derived from CycleGAN introduces artifacts in the results, which is the common problem of CycleGAN-based models. STST (d) derived from cGAN is not able to preserve the correct stain style.
Since our RestainNet uses the $L$ channel from the $Lab$ color space as the model input, it can greatly preserve the structural information. Defining the digital dyes in every pixel can ensure the correctness of the stain style. Furthermore, self-supervised learning alleviates the requirements of paired training data.

\subsection{Stain Normalization in the Segmentation Task}
\label{sec:seg}
\begin{table}
	\caption{Quantitative comparisons of different normalization methods in the segmentation task. Ori is the original images without stain normalization preprocessing.}
	\begin{center}
		\begin{tabular}{lcccc}
			\hline
			Method & Dice & ACC & Pre & Recall\\
			\hline
			(1) Ori     							  &0.842&0.844&0.745&0.773\\
			(2) Reinhard~\cite{reinhard2001color}	  &0.856&0.873&0.872&0.824\\
			(3) Macenko~\cite{2009A}				  &0.861&0.876&0.878&0.833\\
			(4) Vahadane~\cite{vahadane2016structure} &0.856&0.872&0.874&0.840\\
			(5) StainGAN~\cite{Shaban2019Staingan}	  &0.872&0.880&0.859&0.852\\
			(6) STST~\cite{salehi2020pix2pix}		  &0.859&0.866&0.851&0.853\\
			(7) \textbf{Ours}						  &\textbf{0.878}&\textbf{0.883}&\textbf{0.883}&\textbf{0.861}\\
			\hline
		\end{tabular}
	\end{center}
	\label{tab:segmentation}
\end{table}

\begin{figure}[t]
	\setlength{\tabcolsep}{1pt}
	\begin{tabular}{lcccc}
		A&
		\includegraphics[width=.23\linewidth]{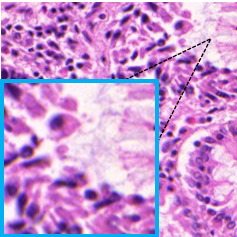}&
		\includegraphics[width=.23\linewidth]{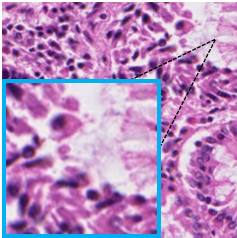}&
		\includegraphics[width=.23\linewidth]{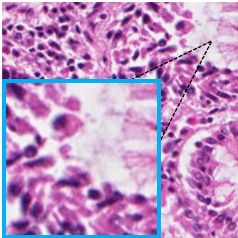}&
		\includegraphics[width=.23\linewidth]{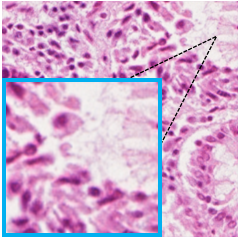}\\
		B&
		\includegraphics[width=.23\linewidth]{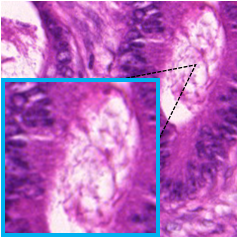}&
		\includegraphics[width=.23\linewidth]{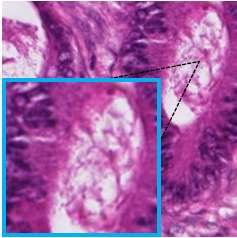}&
		\includegraphics[width=.23\linewidth]{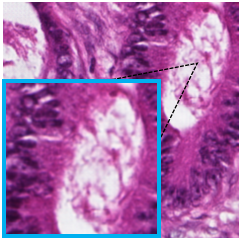}&
		\includegraphics[width=.23\linewidth]{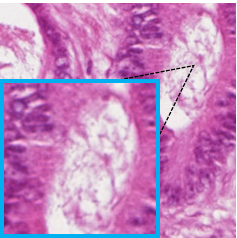}\\
		C&
		\includegraphics[width=.23\linewidth]{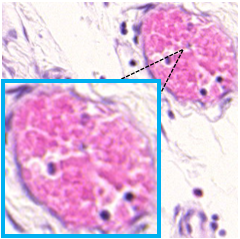}&
		\includegraphics[width=.23\linewidth]{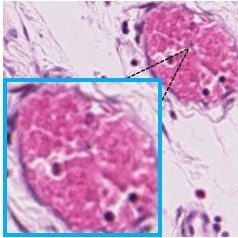}&
		\includegraphics[width=.23\linewidth]{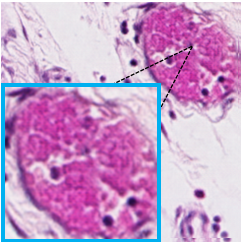}&
		\includegraphics[width=.23\linewidth]{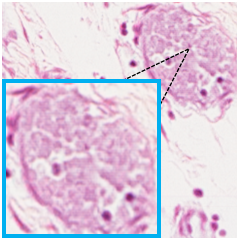}\\
		D&
		\includegraphics[width=.23\linewidth]{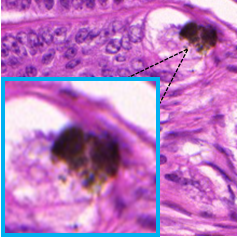}&
		\includegraphics[width=.23\linewidth]{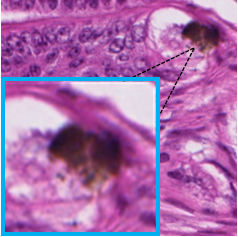}&
		\includegraphics[width=.23\linewidth]{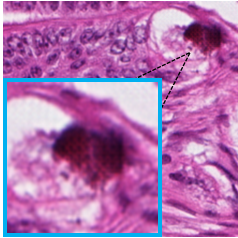}&
		\includegraphics[width=.23\linewidth]{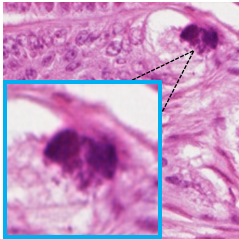}\\
		&(a) Input & (b) Ours & (c) StainGAN & (d) STST
	\end{tabular}
	\caption{Comparisons of stain normalization in MICCAI’16 GlaS challenge dataset. Our proposed RestainNet successfully preserves the structural information with richer and sharper textures.}
	\label{fig:diff_computational_methods}
\end{figure}

In this experiment, we directly applied the stain normalization algorithms on a MICCAI’16 GlaS challenge dataset~\cite{2016Gland} as the preprocessing step. To be fair, the training sets with different stain normalization models were trained with the same network architecture U-net~\cite{2015U} with the same epochs.
All learning-based stain normalization methods were trained on the MITOS-ATYPLA14 dataset in this experiment. 
The quantitative results of the segmentation under different stain normalization methods are shown in Table~\ref{tab:segmentation}. 

We can easily observe that all the stain normalization methods can reduce the stain inconsistency of the data, thereby help improve the segmentation task, shown in (2) to (7). 
(4) uses the non-negative matrix factorization method to obtain the dye concentration matrix to generate a more detailed normalized image, so obtains a better segmentation effect than (2) and (3). On the contrary, (4) also consumes a longer time in stain normalization.
In this task, the segmentation model maps pixels into differentiated representations based on the significant color and morphological differences between tissues to achieve the purpose of gland segmentation. Therefore, to ensure the efficiency of the segmentation model, the normalized image should have a broader color distribution to maintain the distinguishability of the histological tissue clusters.
(6) to (7) implement the conversion between different styles of pathological images through a learning-based way, which can get normalized images with richer colors because they have learned a large number of different types of tissues in the training phase. 
Although the same dataset is employed, the motivations in (5) to (7) are different, resulting in a visible difference in stain style in the synthesized images. (5) uses semi-supervised to achieve the transfer between two image styles in the training set and shows satisfactory performance in the MITOS-ATYPLA14 database, but when predicting on the unfamiliar MICCAI’16 GlaS dataset, the performance weakened. 
(6) also employs an semi-supervised method to color the grayscale images, so it may be difficult to handle the tissue distribution that does not exist in the training data.

To further compare the performance of the learning-based normalization methods, we show normalized results of MICCAI’16 GlaS dataset in Fig.~\ref{fig:diff_computational_methods}. 
We can find that the image generated by our proposed model in (b) has richer and sharper textures than the other two models, as shown in Fig.~\ref{fig:diff_computational_methods}-A and Fig.~\ref{fig:diff_computational_methods}-B.
This is because RestainNet refers to the dye concentration components rather than just the texture information, which reduces the influence of the background area and forces the network to be sensitive to dyes.
Thanks to the dye concentration matrix, the proposed network is also sensitive to the color of the tissue clusters. We can find that (d) shows insufficient tissue recognition ability and colors all tissues pink. Since the blood cells in Fig.~\ref{fig:diff_computational_methods}-C are not combine with the dyes, they should show the original red color of hemoglobin, while (c) and (d) transfer it to purple. 
In Fig.~\ref{fig:diff_computational_methods}-D, the contamination should maintain the original color, but (c) and (d) transfer it as the normal tissue.
These all highlight the benefits of the dye concentration matrix to our model.

\subsection{Stain Normalization in the Classification Task}
\label{sec:class}

\begin{table}[t]
	\caption{Quantitative comparisons of different normalization methods in the classification task. Ori is the original images without stain normalization preprocessing.}
	\centering
	\begin{tabular}{lccc}
		\hline
		Method & F1-score & Accuracy & Precision \\
		\hline
		(1) Ori     							&0.815&0.914&0.818\\
		(2) Reinhard~\cite{reinhard2001color}	&0.903&0.930&0.906\\
		(3) Macenko~\cite{2009A}				&0.895&0.923&0.896\\
		(4) StainGAN~\cite{Shaban2019Staingan}	&0.905&0.931&0.911\\
		(5) STST~\cite{salehi2020pix2pix}		&0.879&0.912&0.886\\
		(6) \textbf{Ours}						&\textbf{0.910}&\textbf{0.935}&\textbf{0.919}\\
		\hline
	\end{tabular}
	\label{tab:classification}
\end{table}

In this experiment, we conduct a classification task on Zhao dataset~\cite{0Artificial} to examine the influence of various normalization algorithms on the classification task. 
The classification task in this experiment is more challenging than the segmentation task described in Sec.~\ref{sec:seg} for the following two reasons. First, in the Zhao dataset, the data obtained by multiple centers brings serious inconsistencies in stain styles, which requires the normalization algorithm to have a better generalization.
Secondly, there are night types of tissues with different dye adsorption capabilities in the Zhao dataset, which poses the normalization algorithm with a great challenge to identify inconsistent stained areas.
Similar to Section~\ref{sec:seg}, we first use the normalization algorithm for preprocessing then construct the standard ResNet34 network to classify all the tissues. Except that the stain normalization method is different, the structure, the training process, and the testing process of the model are the same. The quantitative results are described in Table~\ref{tab:classification}.

Since the data were collected from different institutions, there are significant stain styles differences due to different scanners and pathological section production specifications, which harm the performance of the classification model. As shown in Table~\ref{tab:classification}, we can observe different degrees of performance improvement of the classification task with different stain normalization methods. Our proposed model brings the most significant benefit for the classification task among the five approaches, which demonstrates the effectiveness of our proposed method.

\subsection{Model Stability}
\label{sec:stability}
\begin{figure*}[t]
	\centering
	\includegraphics[width=\linewidth]{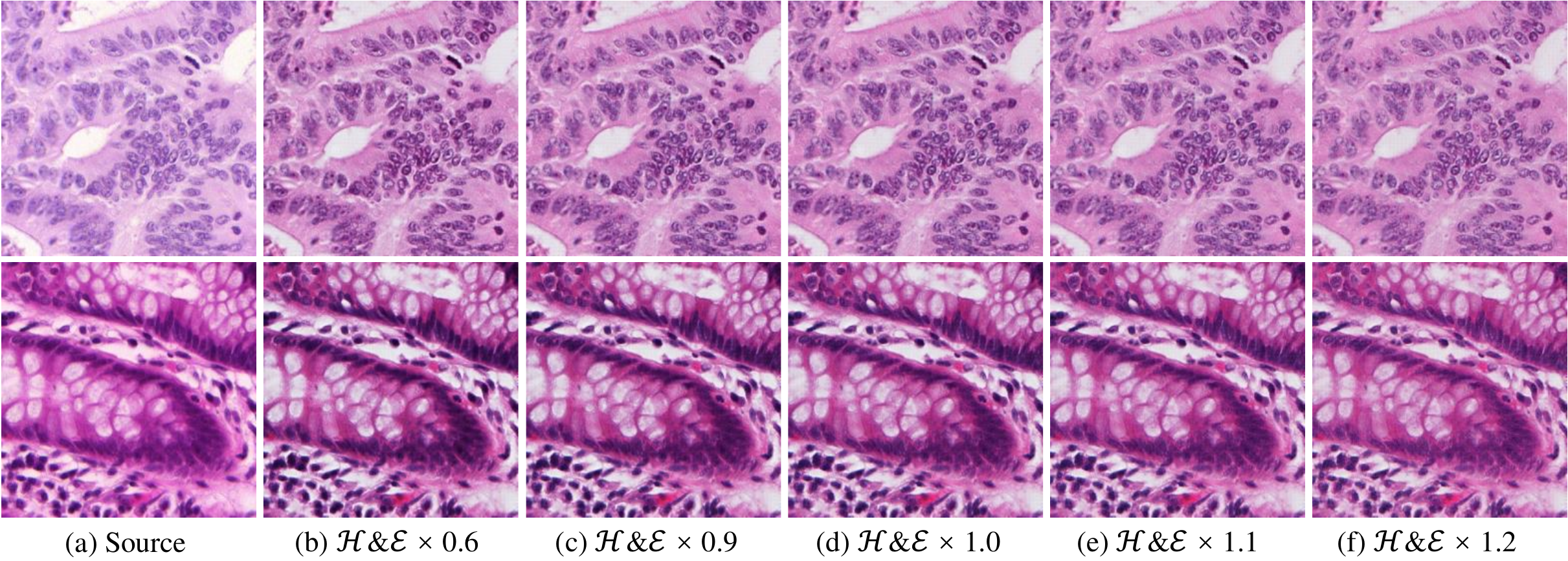}
	\caption{(a) is the input images without stain normalization. (b) to (f) indicate the outputs of the RestainNet of with different vibration of $\mathcal{H}$ and $\mathcal{E}$ components by different multiplication coefficients 0.6, 0.9, 1.0, 1.1, and 1.2, respectively.}
	\label{fig:re_nor}
\end{figure*}

We have already demonstrated the superiority of our proposed model in the previous experiments. In this part, we want to evaluate the stability of the proposed RestainNet, to see how the input $\mathcal{H}$ and $\mathcal{E}$ really affect the network outputs. Because even $\mathcal{H}$ and $\mathcal{E}$ components extracted by the Beer-Lambert's Law have demonstrated a certain degree of tolerance in stain style inconsistency. It may still be various when using different staining protocols or scanners. As an crucial part of the network input, we discuss how $\mathcal{H}$ and $\mathcal{E}$ work in the network. The this end, we add some vibrations to $\mathcal{H}$ and $\mathcal{E}$ by manually introducing a set of multiplication coefficients $\{0.6, 0.9, 1.0, 1.1, 1.2\}$.

Fig.~\ref{fig:re_nor} demonstrates the results with the input $\mathcal{H}$ and $\mathcal{E}$ multiplied by coefficients. We can find that the staining style of the results from Fig.~\ref{fig:re_nor} (b)-(f) is actually preserved very well even with different degrees of vibrations. This observation proves that our proposed model has successfully learn the staining style of the source domain. $\mathcal{H}$ and $\mathcal{E}$ deliver the hints to decide whether a pixel should be stained by hematoxylin or eosin, not the staining style. This property is meaningful for the robustness and the stability of the model. 

\subsection{Stain Intensity Preservation}
\label{sec:preservation}

\begin{figure*}[t]
	\setlength{\tabcolsep}{1pt}
	\begin{tabular}{lllllll}
		&
		\includegraphics[width=.16\linewidth]{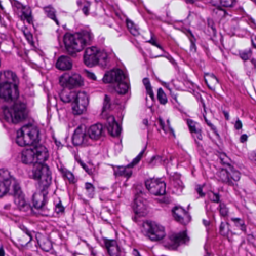}&
		\includegraphics[width=.16\linewidth]{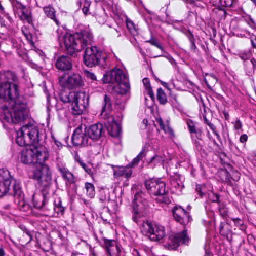}&
		\includegraphics[width=.16\linewidth]{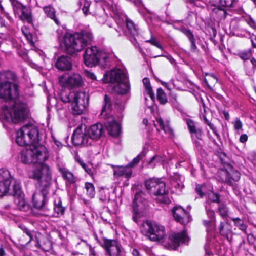}&
		\includegraphics[width=.16\linewidth]{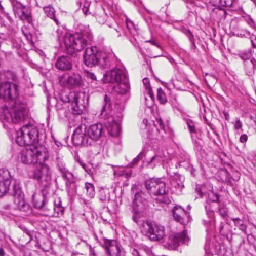}&	
		\includegraphics[width=.16\linewidth]{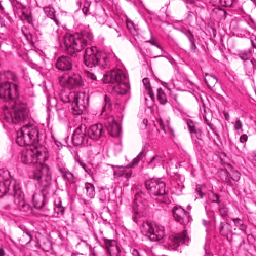}&		
		\includegraphics[width=.16\linewidth]{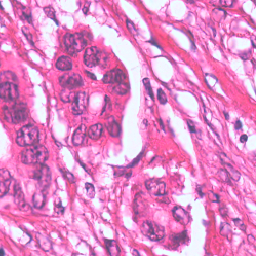}\\
		$\mathcal{H}$ &
		\includegraphics[width=.16\linewidth]{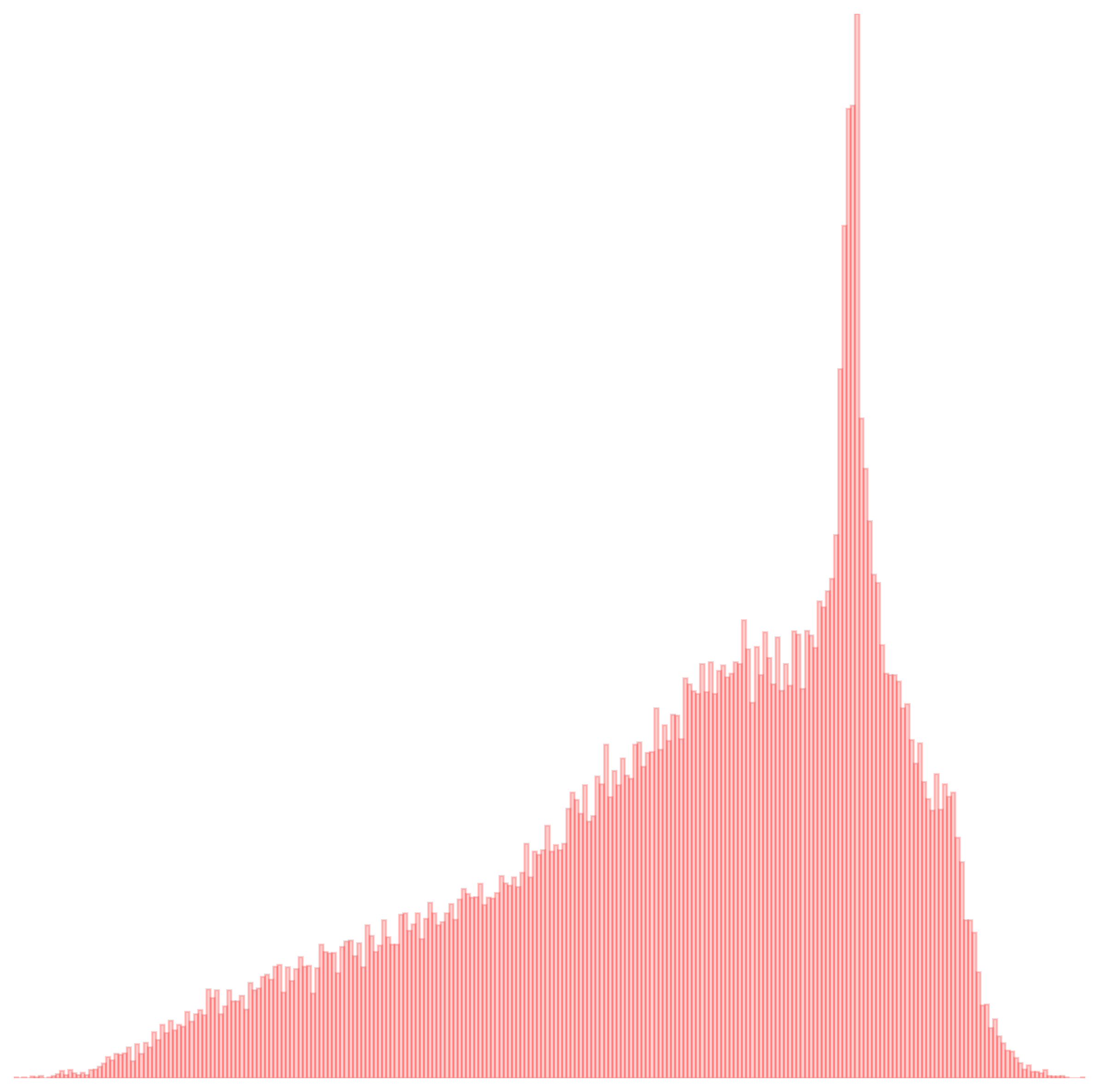}&
		\includegraphics[width=.16\linewidth]{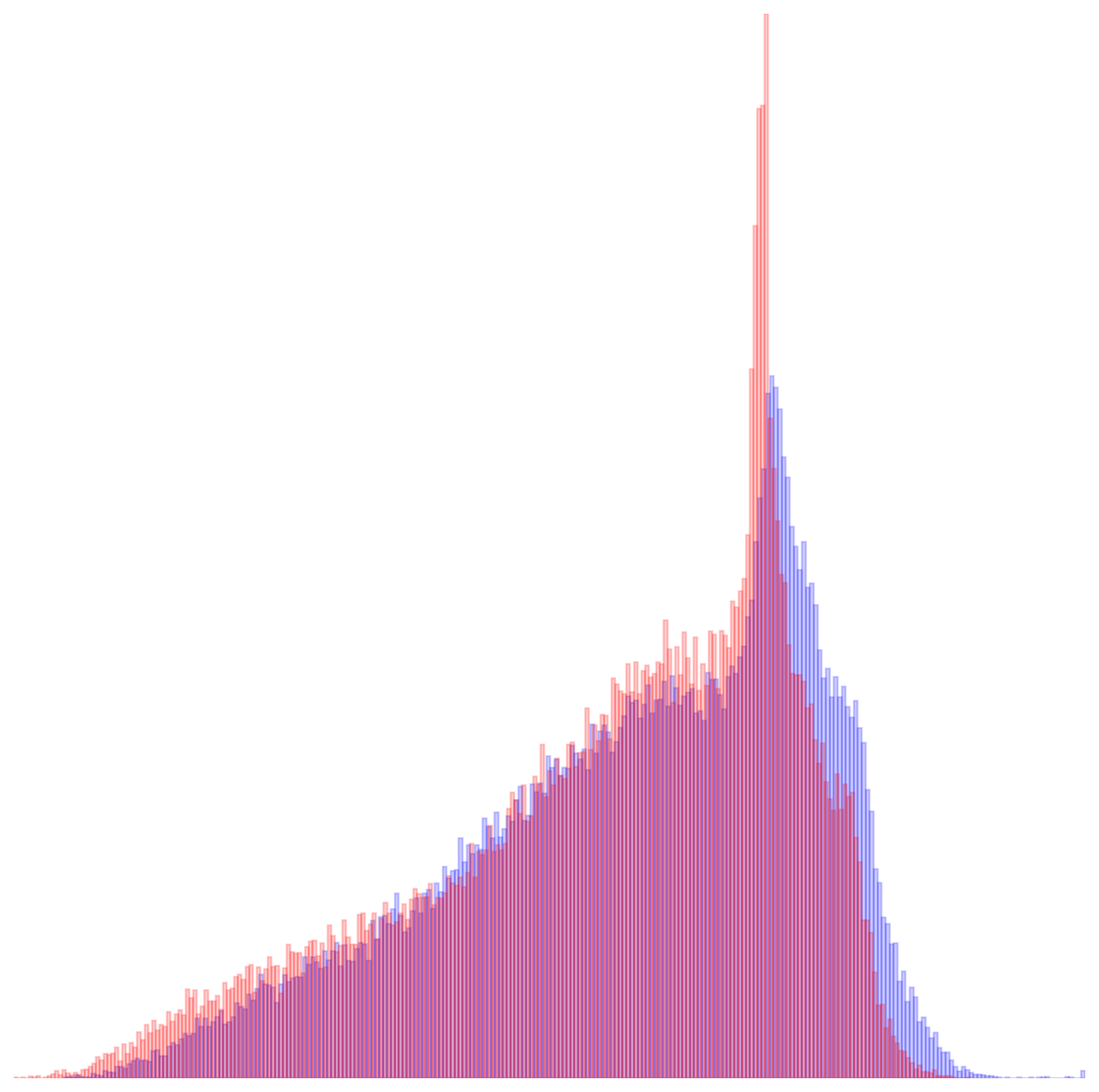}&
		\includegraphics[width=.16\linewidth]{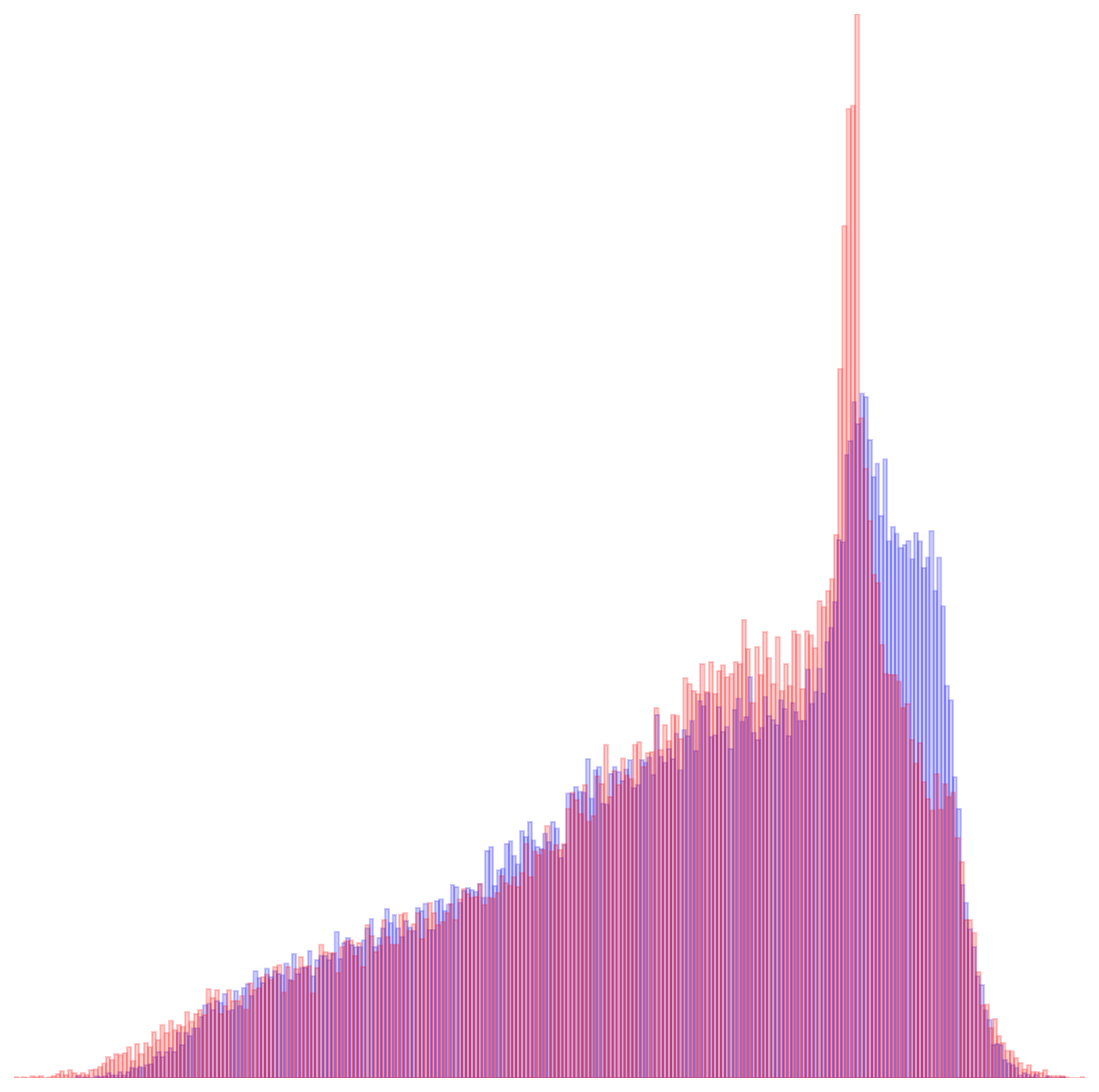}&
		\includegraphics[width=.16\linewidth]{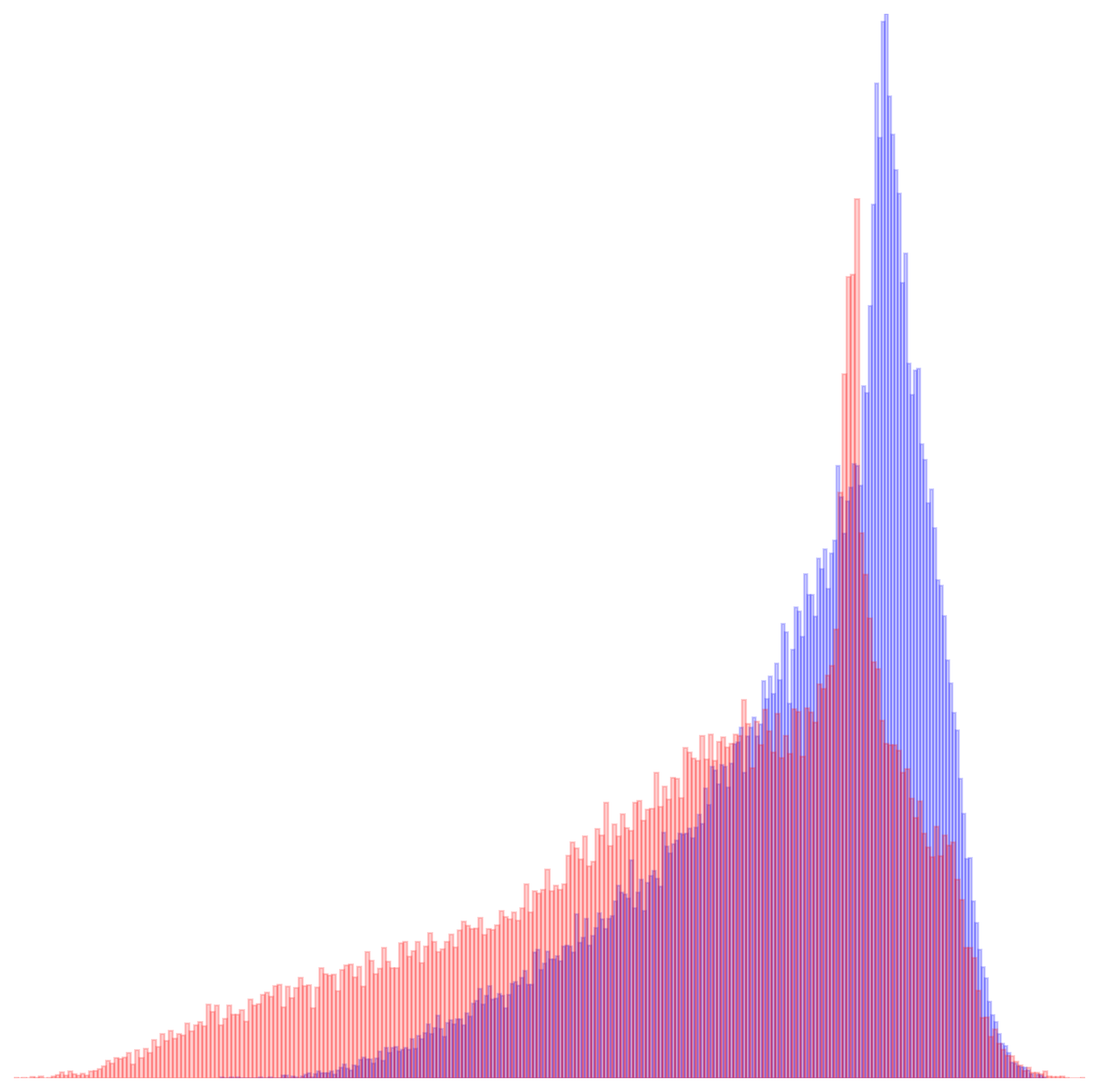}&
		\includegraphics[width=.16\linewidth]{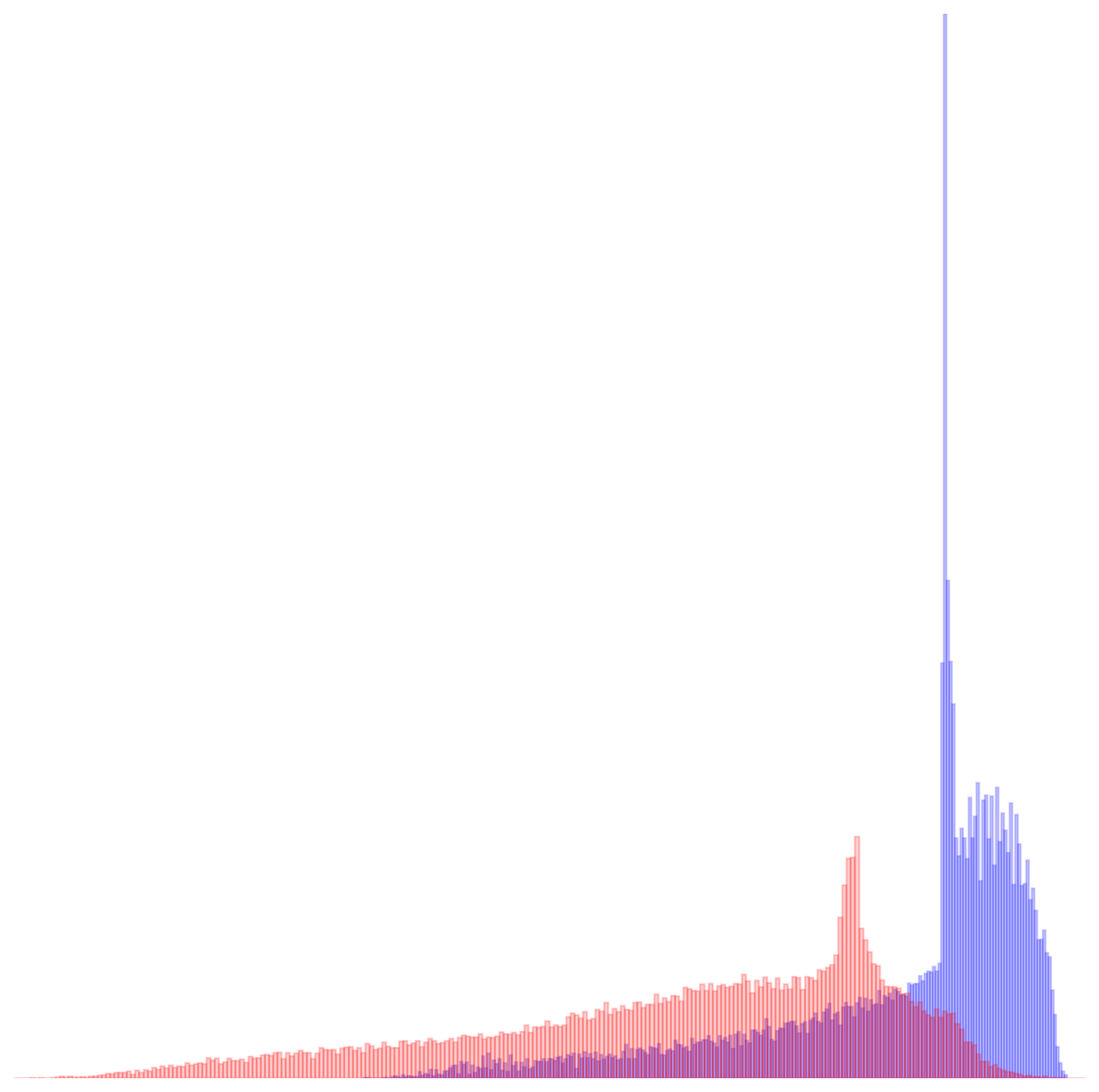}&
		\includegraphics[width=.16\linewidth]{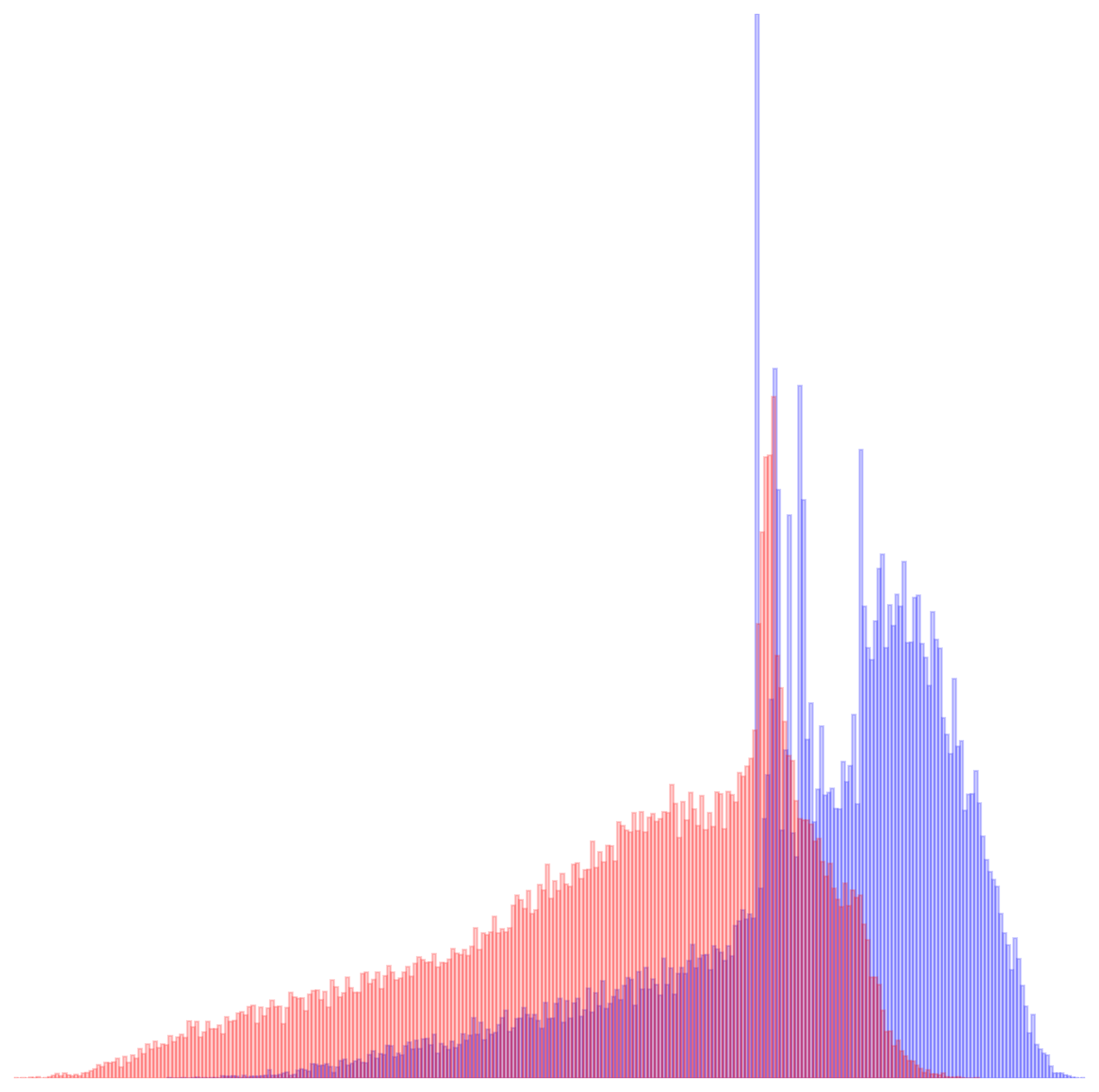}\\
		$\mathcal{E}$ &
		\includegraphics[width=.16\linewidth]{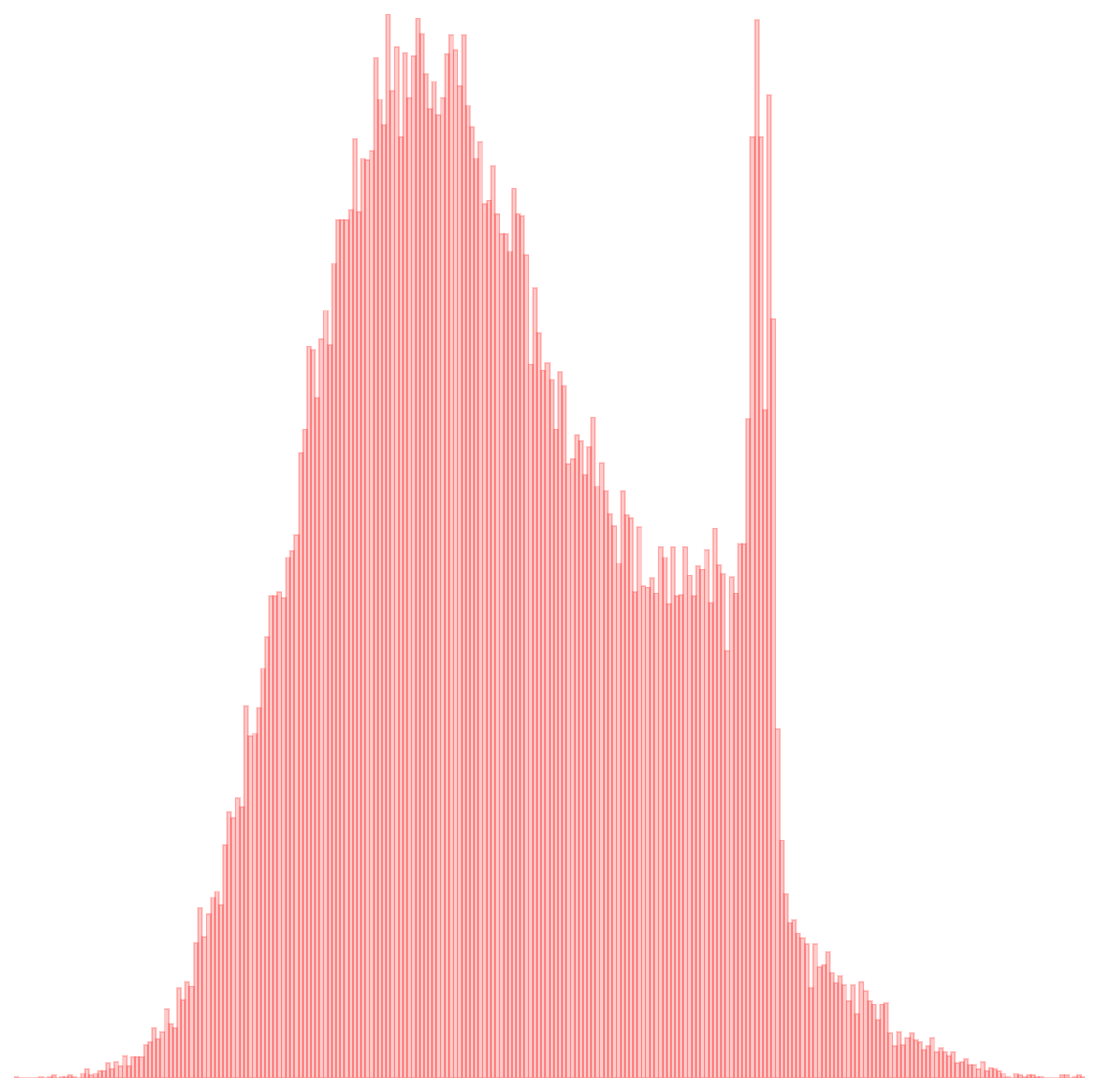}&
		\includegraphics[width=.16\linewidth]{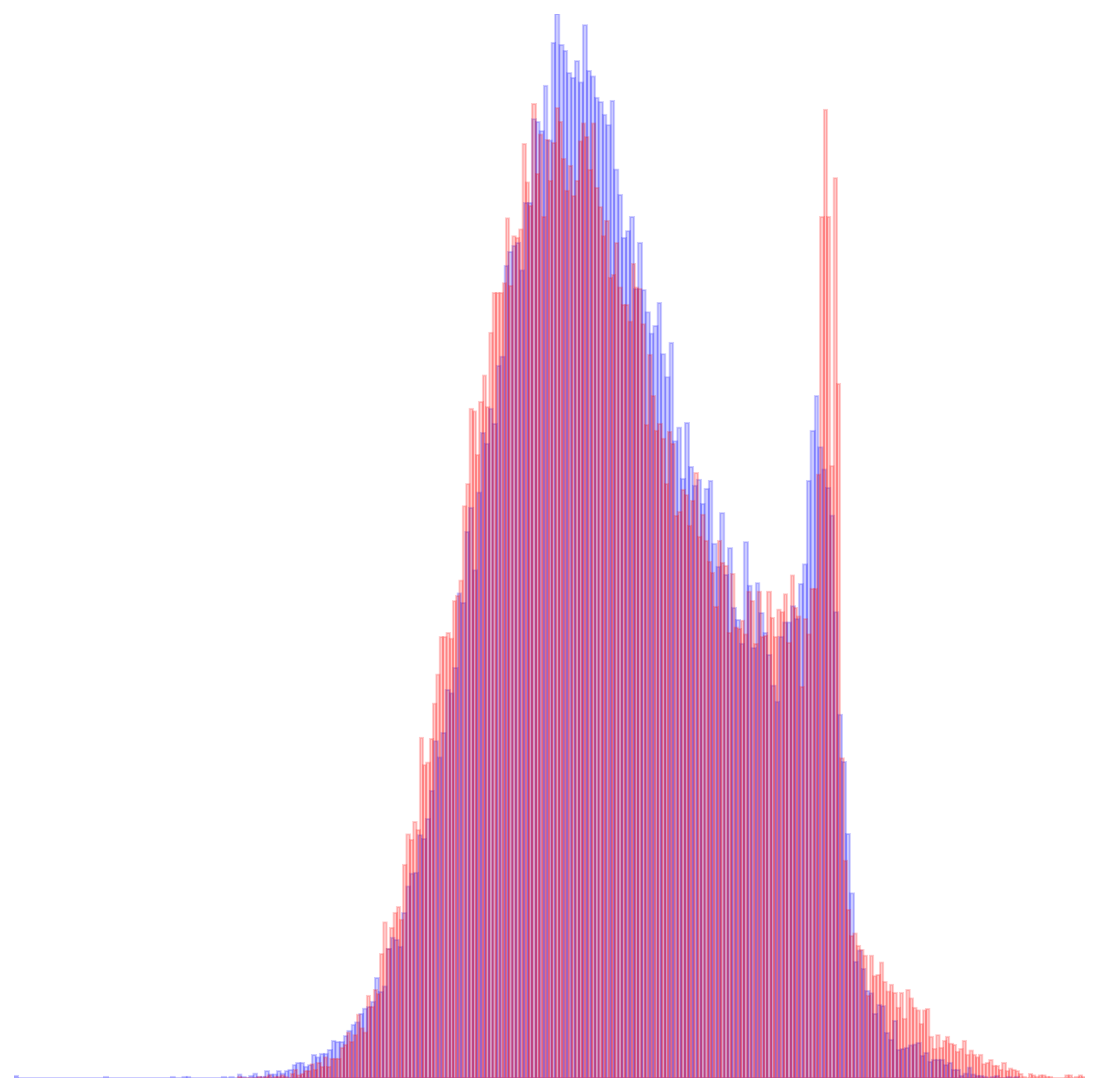}&
		\includegraphics[width=.16\linewidth]{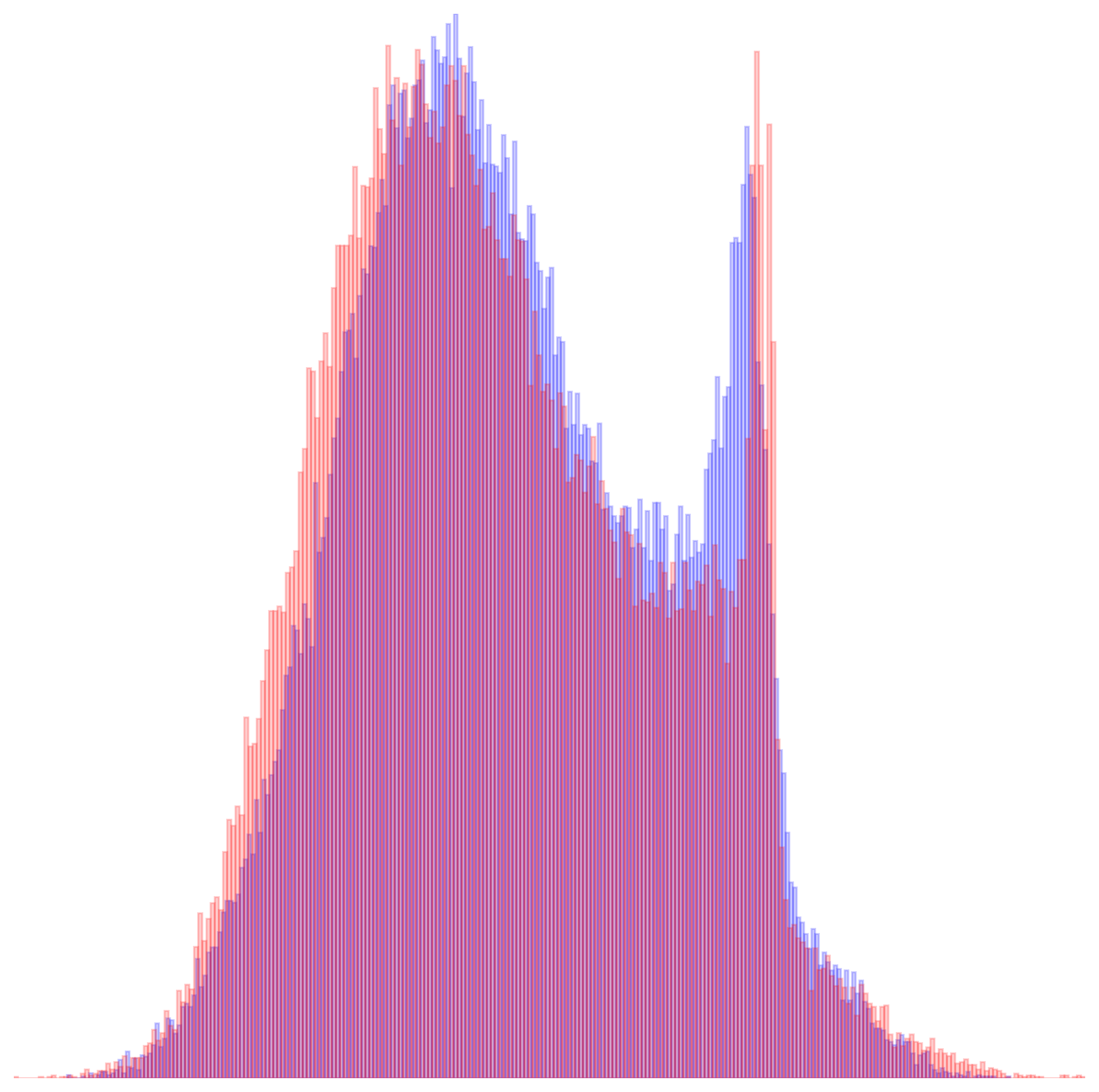}&
		\includegraphics[width=.16\linewidth]{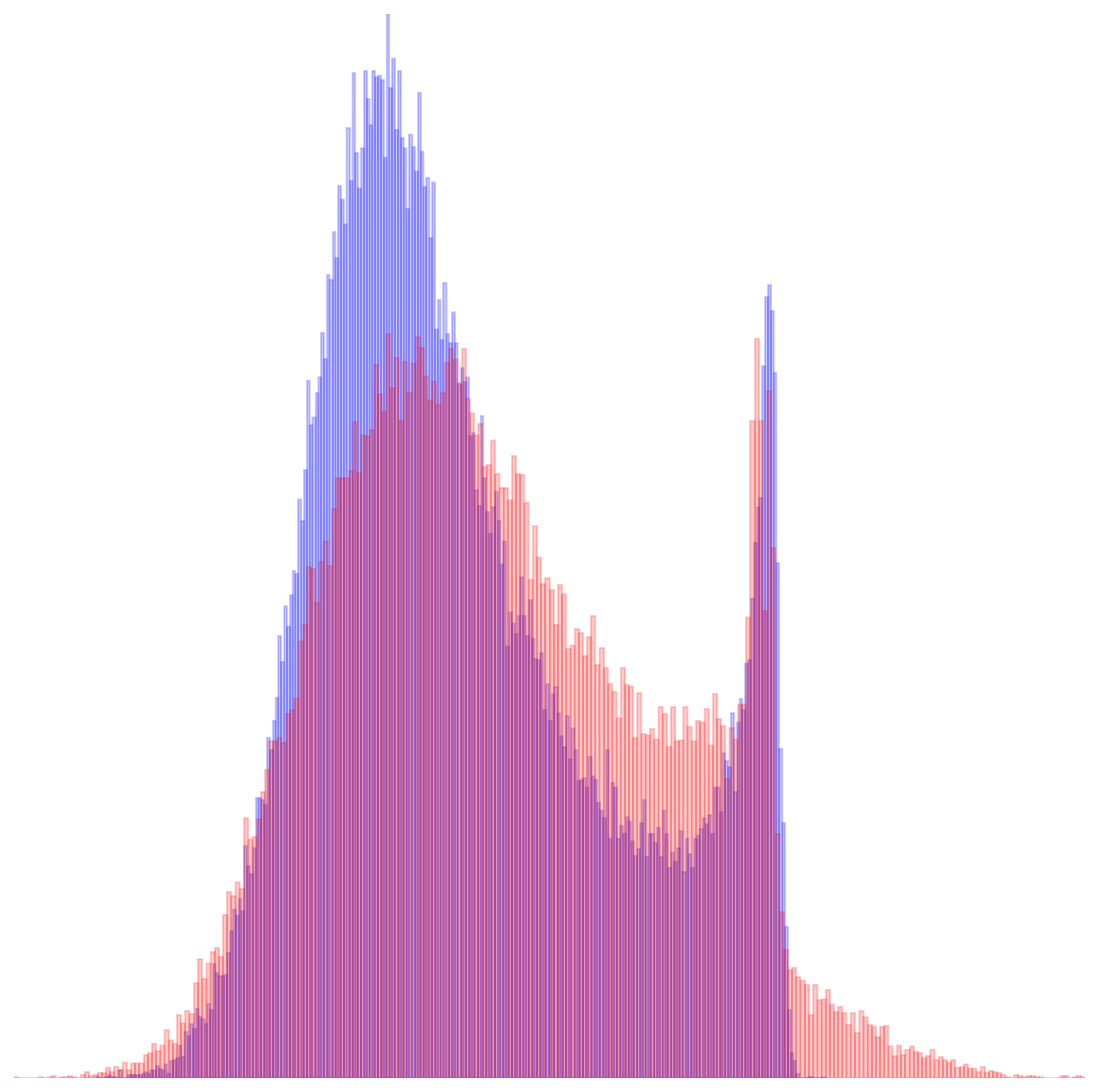}&
		\includegraphics[width=.16\linewidth]{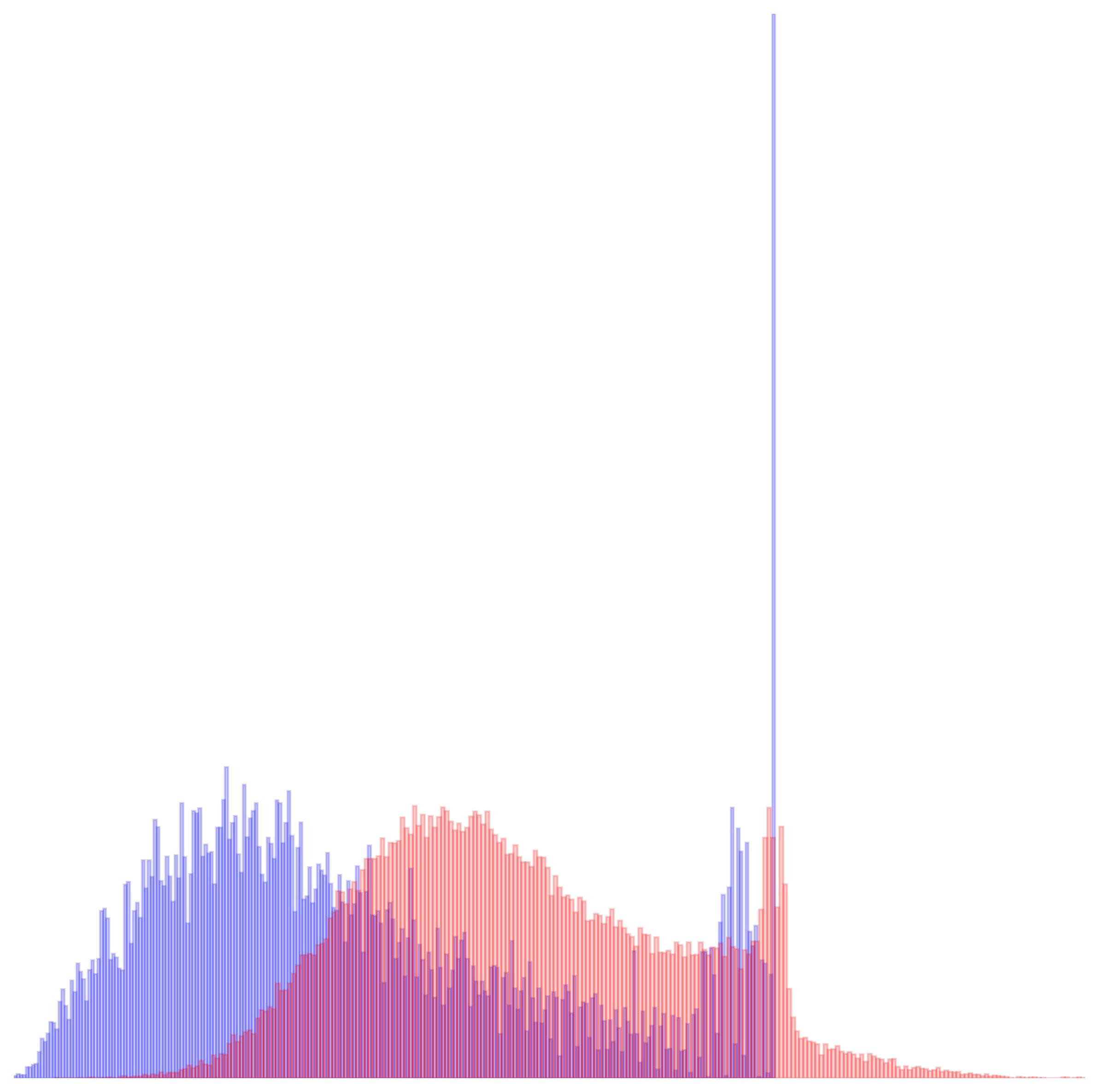}&
		\includegraphics[width=.16\linewidth]{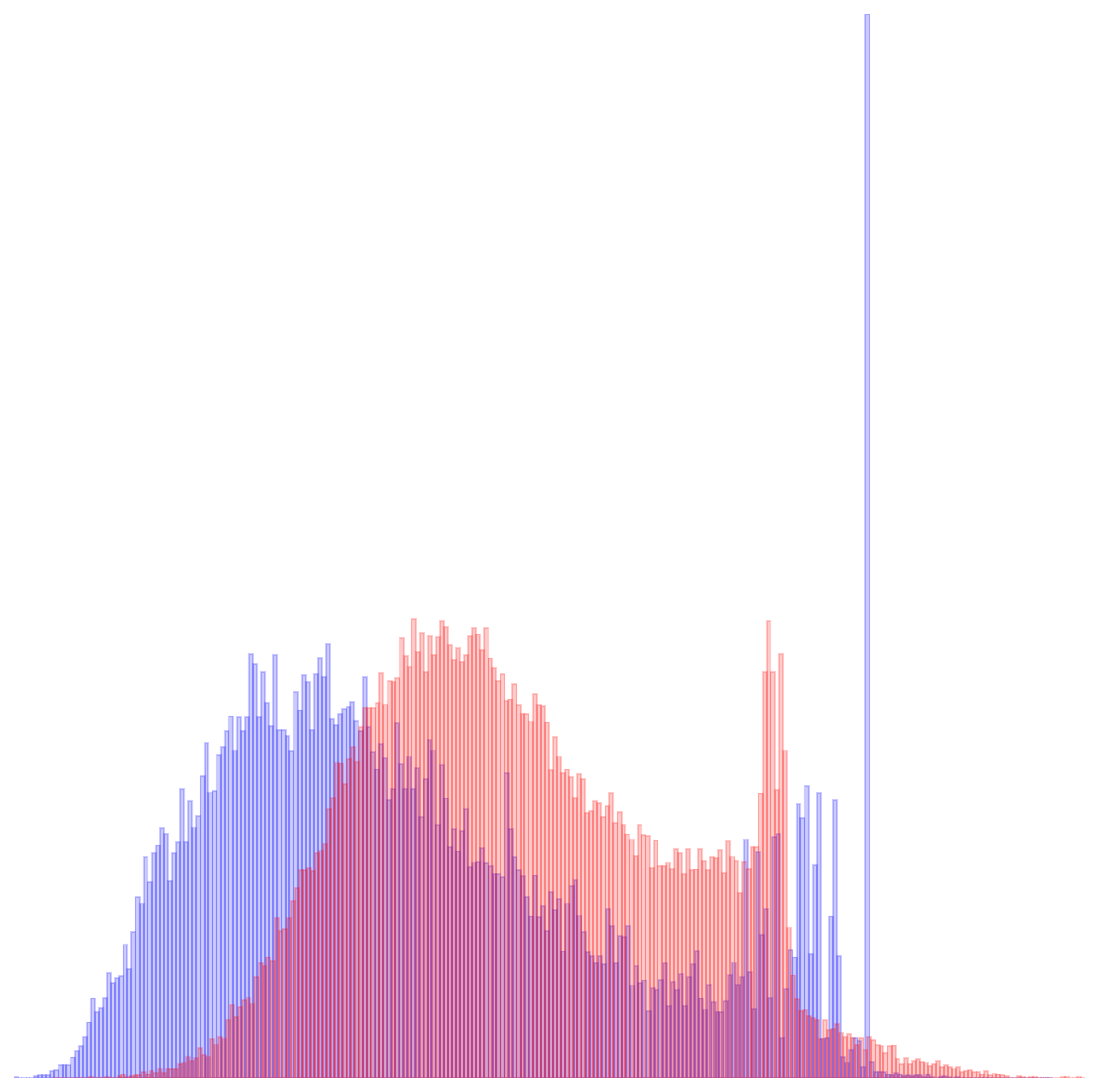}\\
		& (a) GT & (b) Ours & (c) StainGAN  & (d) STST & (e) Vahadane & (f) Reinhard
	\end{tabular}
	\caption{Dye intensity preservation. GT are the images from the target domain (scanned by Hamamatsu scanner). $\mathcal{H}$ and $\mathcal{E}$ are the histograms of dye intensity. We overlay the histogram of the generated image (blue) on the ground truth (pink). Dye intensity is well preserved by our model.}
	\label{fig:hist}
\end{figure*}

The results in Table~\ref{tab:result} show the excellent performance of our model in protecting the structure of the image. Not only that, but the normalization algorithm should also preserve the dye component to maintain the optical characteristics. 
To this end, we compare the gray histogram of the normalized image and the input to qualitatively evaluate the ability of the algorithm to protect the dye components. 
As shown in Fig.~\ref{fig:hist}, the first row is the RGB image, and the second and third rows are the gray histograms of $\mathcal{H}$ and $\mathcal{E}$, respectively. 

In general, the method involved in this paper can maintain the texture structure of the image, but the stain style of the generated image is different.
We observe that the computational-based normalization methods (e.g. (c) and (d)) are template dependent which can only generate the selected color style, and their grayscale histograms are the most different from (a). 
(d) uses the same training data as (b) to color the grayscale images, but discard the color information of the input image, so it is difficult to distinguish the tissue components, resulting in inefficient normalization. 
(c) consumes more computing resources to construct a more powerful generative network and synthesize a normalized image highly similar to (a).  
However, (c) does not obtain prior knowledge of the digital dyes, which leads to the error coloring of the pathological tissue, harming the consistency of the histogram with (a). 
Thanks to the re-stain mechanism and customized staining loss, our model can synthesize the normalized image that best preserving the dye components, thereby promoting the performance of the subsequent models (e.g. segmentation or classification model), as described in the Sec~\ref{sec:class} and Sec~\ref{sec:seg}.

\begin{figure}[t]
	\centering
	\setlength{\tabcolsep}{1pt}
	\begin{tabular}{cccc}
		\includegraphics[width=.33\linewidth]{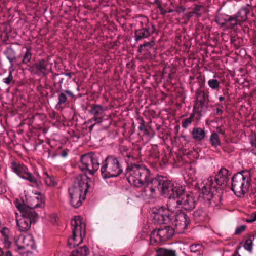}&
		\includegraphics[width=.33\linewidth]{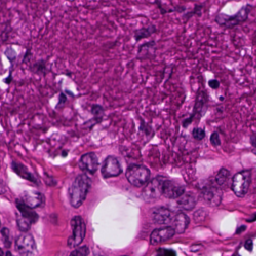}&
		\includegraphics[width=.33\linewidth]{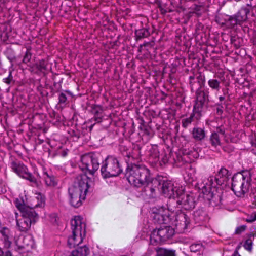}\\
		(a) Aperio & (b) Hamamatsu & (c) Ours
	\end{tabular}
	\caption{A challenging case of the over-stained images.}
	\label{fig:shortcome}
\end{figure}
\section{Conclusion}
\label{sec:Conclusion}
In this work, we formulated stain normalization in a new perspective which regarded it as a digital staining process. To achieve this goal, we present a self-supervised model RestainNet. Besides superior quantitative and qualitative results, our model demonstrates the greatest flexibility in practical usage comparing with existing approaches. First of all, RestainNet does not require paired data for network training thanks to the self-supervised learning nature. Second, since RestainNet learns the color distribution of one specific domain (scanner) in a self-supervised manner. It allows transferring input pathology images from any other domains to the target domain without additional training cost.

However, there is still one challenge we have to tackle in future works. Since our model learns the color distribution of the specific domain. It mainly focuses on solving the color inconsistency problem introduced by different scanners. When color inconsistency is introduced by human error during sectioning, it will become a challenge of our model. Fig.~\ref{fig:shortcome} demonstrates an extreme case which is an over-stained section scanned by two different scanners. Our RestainNet achieves state-of-the-art performance comparing with existing models in three different tasks, the color normalization task, the segmentation task and the classification task.

%
\bibliography{reference}
\bibliographystyle{IEEEtran}

\end{document}